\documentclass[letterpaper, 10pt, conference]{ieeeconf}      

\IEEEoverridecommandlockouts                              
                                                          
\usepackage{times}
\usepackage{amsmath}
\usepackage{mathrsfs}
\usepackage{amssymb}
\usepackage{scalerel}
\usepackage{mathtools}
\usepackage{xcolor}
\usepackage{kotex}
\usepackage{cite}
\usepackage{graphicx}
\usepackage{algorithm}
\usepackage{algorithmic}
\usepackage{comment}
\usepackage{adjustbox}
\usepackage{romannum}
\usepackage[inkscapeformat=png]{svg}
\usepackage{multirow}
\usepackage{tabularx}
\usepackage{siunitx}
\usepackage{tikz}
\usepackage{tikz-cd}
\usepackage{pgfplots}
\usepackage{grffile}
\usepackage{tikzscale}
\usepackage{lipsum}
\usepackage{verbatim}
\usepackage{setspace}
\usepackage{cuted}
\usepackage{breqn}
\usepackage{makecell}
\usepackage[dvipsnames]{xcolor}
\usepackage{bm}
\let\labelindent\relax
\usepackage{enumitem}
\usepackage{url}
\usepackage{cancel}

\definecolor{darkblue}{rgb}{0,0,0.65}

\pgfplotsset{compat=newest}
\usetikzlibrary{plotmarks}
\usetikzlibrary{arrows.meta}
\usetikzlibrary{shapes,arrows}
\usetikzlibrary{arrows,calc,positioning}
\usepgfplotslibrary{patchplots}
\usetikzlibrary{decorations.pathreplacing}

\usepackage{pgfplots}
\pgfplotsset{compat=1.18}

\pgfplotsset{
  every axis/.append style={
    scale only axis,
    enlargelimits=false,
    tick label style={font=\normalsize},
    label style={font=\normalsize},
    title style={font=\normalsize},
  }
}

\pgfplotsset{
  table/search path={figure/lienard_constant}
}

\usepackage{amsthm}

\newtheorem{thm1}{\bf Theorem}
\newtheorem{prop1}{\bf Proposition}
\newtheorem{lem1}{\bf Lemma}
\newtheorem{asm1}{\bf Assumption}
\newtheorem{defn1}{\bf Definition}
\newtheorem{rmk1}{\bf Remark}
\newtheorem{exmp1}{\bf Example}
\newtheorem{cor1}{\bf Corollary}

\newtheorem{prob1}{\bf Problem}
\newtheorem{sublem}{Lemma} 

\newcommand{\scriptA}{\mathcal{A}}
\newcommand{\scriptB}{\mathcal{B}}
\newcommand{\scriptC}{\mathcal{C}}
\newcommand{\scriptD}{\mathcal{D}}
\newcommand{\scriptE}{\mathcal{E}}
\newcommand{\scriptF}{\mathcal{F}}
\newcommand{\scriptG}{\mathcal{G}}
\newcommand{\scriptH}{\mathcal{H}}
\newcommand{\scriptI}{\mathcal{I}}

\newcommand{\scriptK}{\mathcal{K}}
\newcommand{\scriptL}{\mathcal{L}}
\newcommand{\scriptM}{\mathcal{M}}
\newcommand{\scriptN}{\mathcal{N}}
\newcommand{\scriptO}{\mathcal{O}}
\newcommand{\scriptP}{\mathcal{P}}

\newcommand{\scriptS}{\mathcal{S}}
\newcommand{\scriptT}{\mathcal{T}}

\newcommand{\frakturB}{\mathfrak{B}}

\newcommand{\frakturS}{\mathfrak{S}}

\newcommand{\doubleE}{\mathbb{E}}

\newcommand{\doubleN}{\mathbb{N}}

\newcommand{\doubleR}{\mathbb{R}}

\newcommand{\doubleZ}{\mathbb{Z}}

\newcommand{\product}[3]{\prod_{#1}^{#2} #3}

\newcommand{\defeq}{\vcentcolon=}
\newcommand{\eqdef}{=\vcentcolon}

\newcommand{\longbar}[1]{\longbar{#1}}

\allowdisplaybreaks[4]

\title{\LARGE \bf A Distributed Observer with Random Agent Activations over Random Communication Networks}

\author{Yong Joo Do and Jin Gyu Lee 
\thanks{*This work was supported by the National Research Foundation of Korea (NRF) grant funded by the Korea government (MSIT) (RS-2026-25504174).
}
\thanks{All authors are with ASRI, Department of Electrical and Computer Engineering, Seoul National University, 1 Gwanak-ro, Gwanak-gu, Seoul, 08826, Korea. Email: {\tt\small yjdo@cdsl.kr}; {\tt\small jingyu.lee@snu.ac.kr}}%
}

\begin{document}

\maketitle
\thispagestyle{empty}
\pagestyle{empty}

\begin{abstract}
    This paper develops a distributed observer for the forward-Euler discretization of a continuous-time linear time-invariant system under random local operations and random communications. Under uniform stochastic strong connectivity, we establish almost sure exponential convergence of all local state estimates for sufficiently small discretization step sizes. We further examine how the underlying parameters and components affect the admissible range of the discretization step size and the guaranteed convergence rate bound. Unlike related approaches, the proposed method does not require each local unobservable subspace to admit an $A$-invariant complement. 
\end{abstract}


\section{Introduction}\label{sec:introduction}
    
    Given a continuous-time linear time-invariant (LTI) system 
    \vspace{-9pt}
    \begin{equation}\label{eqn:conti}
        \dot{x}(t) = Ax(t),
        \vspace{-3pt}
    \end{equation}
    where $x(t)\in\doubleR^{n}$ denotes the state at time $t$, suppose that the model \eqref{eqn:conti} is implemented on a digital computing platform. We study the cooperative reconstruction of the state of this digital implementation by agents in $\scriptN \defeq \{1,\ldots,N\}$. Each agent has access only to its assigned local output, which is generated directly from that state. Specifically, we consider
    \vspace{-3pt}
    \begin{equation}\label{eqn:discretized}
        x[k+1] = \Phi_{\alpha} x[k],\quad y_{i}[k] = C_{i}x[k],\quad i \in \scriptN,
        \vspace{-3pt}
    \end{equation}
    where $x[k]\in\doubleR^{n}$ denotes the internal state of the digital implementation, $y_{i}[k]\in\doubleR^{p_{i}}$ denotes the local output available to agent $i$, and $\Phi_{\alpha}$ is the transition matrix associated with discretization. The discretization step size $\alpha$ is identified with the sampling period of the digital implementation. The goal of each agent is to reconstruct $x[k]$ using its local output and information exchanged with neighboring agents. 
    
    Several discretization schemes can be used to obtain \eqref{eqn:discretized}. In particular, the exact discretization $\Phi_{\alpha}=e^{\alpha A}$ reproduces the continuous-time state at the sampling instants. For simplicity, however, we adopt the forward-Euler model $\Phi_{\alpha}=I_{n}+\alpha A$. 

    \vspace{-3pt}
    \subsection{Motivating Example}

        A motivating example is a digitally implemented chaotic communication system, in which the receiver reconstructs the transmitter's internal state from the transmitted output \cite{NM97,SZ17}. In such a scheme, a message $m[k]$ is masked by some chaotic internal state $x[k]$ at the transmitter. The state $x[k]$ is generated by implementing a continuous-time oscillator, such as the R\"ossler system \cite{Ro76}, on a field-programmable gate array (FPGA) using forward-Euler updates \cite{SZ17}. The oscillator model and its parameters are assumed to be known a priori to both the transmitter and receiver \cite{NM97,SZ17}. The receiver uses the output $y[k]$ sent from the transmitter to reconstruct $x[k]$ and recover the original message $m[k]$. However, if a third party (somehow) obtains the oscillator model and its parameters, disclosure of the synchronization output $y[k]$ may enable that party to reconstruct the full state $x[k]$ analogously. This motivates considering a cooperative setting in which multiple receivers obtain different local outputs $y_{i}[k]$ that are individually insufficient, but collectively sufficient, for full-state reconstruction. Each receiver then seeks to reconstruct $x[k]$ using its local output and information exchanged with neighboring receivers. Thus, even if a local output $y_{i}[k]$ is disclosed to a third party, that output alone is insufficient to reconstruct $x[k]$. The above example involves nonlinear dynamics, whereas \eqref{eqn:discretized} is based on a continuous-time LTI model. Nevertheless, the present study serves as a first step toward the framework for such nonlinear settings.

    \vspace{-3pt}
    \subsection{Distributed Observer}

        The cooperative state reconstruction described above is the objective of distributed observers, which have been extensively studied \cite{KLS20}. As noted above, communication among agents plays a central role in this framework. Accordingly, formulating and relaxing communication assumptions are important aspects of distributed observer design and have been a focus of recent research. For example, the observer in \cite{TWF22} uses random pairwise exchanges, while that in \cite{WLM22} allows asynchronous communication and updates. Connectivity assumptions have also been weakened: instantaneous graphs may be disconnected if their union over every prescribed time window is connected \cite{MRBS22,ZLDC24}, or if the network is connected in an average sense \cite{WG24}. The recent work \cite{KLS25} develops a distributed observer accommodating asynchronous local communication over jointly connected switching networks, achieving both asynchronous communications and relaxed communication requirements. Its analysis, however, relies on the restrictive commutativity condition $AP_{i}=P_{i}A$, where $P_{i}$ denotes the orthogonal projector onto the unobservable subspace of $(C_{i},A)$. Motivated by these works, we pose the following question: \emph{can a distributed observer on \eqref{eqn:discretized} be designed, achieving both asynchronous communications and relaxed connectivity requirements without imposing the commutativity condition?}
    \vspace{-3pt}

    \subsection{Main Contribution}
    
        In this paper, we adopt a probabilistic framework with random agent activations and random communication links. Each agent updates its local estimate through \emph{local propagation} at every step and intermittently performs a \emph{projection-averaging} stage as determined by its \emph{activation random variable}; we refer to this mechanism as random activation. The random communication link is assumed to be \emph{uniformly stochastically strongly connected} (Assumption~\ref{asm:USSC}): over each prescribed nonoverlapping time block of fixed length, the union graph is strongly connected with a uniformly positive conditional probability given the past.

        Within this setting, we develop a distributed observer for \eqref{eqn:discretized}. Under joint detectability (Assumption~\ref{asm:jointobs}) and suitable activation and conditional independence conditions  (Assumptions~\ref{asm:activation} and \ref{asm:activation_ind}), we establish almost sure exponential convergence of all local estimates to the plant state for sufficiently small step sizes. Unlike \cite{KLS25}, our analysis does not require the commutativity condition $AP_{i}=P_{i}A$. The key idea is to limit plant evolution per communication opportunity through a sufficiently small step size, allowing the contraction induced by communication and projection-averaging stage to dominate the growth associated with the plant dynamics.

        The paper is organized as follows. Section~\ref{sec:prob} presents the distributed observer under random operations and random communication networks, and states the main convergence result. Section~\ref{sec:prf} proves the main result and discusses the relation to \cite{KLS25}. Section~\ref{sec:simulation} provides the simulation, and Section~\ref{sec:conclusion} concludes the paper.

    \textit{Notation:}
    $\doubleR_{\geq 0}$, $\doubleR_{>0}$, $\doubleZ_{\geq 0}$, and $\doubleN$ denote the sets of nonnegative real numbers, positive real numbers, nonnegative integers, and positive integers, respectively.
    $\|\cdot\|$ denotes the 2-norm for vectors and its induced norm for matrices, while $|\scriptA|$ denotes the cardinality of a set $\scriptA$.
    $\otimes$ denotes the Kronecker product. 
    For $a\in\doubleR$, $\lceil a\rceil$ denotes the smallest integer greater than or equal to $a$. $\operatorname{col}(x_1,\ldots,x_{m})$ denotes the vector obtained by vertically stacking $x_1,\ldots,x_{m}$. $\doubleE[\cdot]$ denotes the expectation, $\mathbf{1}_{\Omega'}$ denotes the indicator function for event $\Omega' \subseteq \Omega$, and $\sigma(\cdot)$ denotes the $\sigma$-algebra generated by its arguments. $\frakturS(F)$ denotes the spectrum of $F$. For any Hurwitz matrix $F$, define its forward-Euler stability threshold as $\alpha_{\mathrm E} (F)\defeq\min_{\lambda\in\frakturS(F)}-2\operatorname{Re}\lambda/|\lambda|^{2}$ \cite[Sec.~IV.2]{Ha96}\footnote{Given a Hurwitz matrix $F$, the forward-Euler discretization $z[k+1]=(I+\alpha F)z[k]$ is exponentially stable if and only if $0<\alpha<\alpha_{\mathrm{E}}(F)$, or equivalently, if and only if $I+\alpha F$ is Schur.}.

\section{Distributed Observer Design}\label{sec:prob}


    We first introduce the local observability decomposition used in the observer design. Let $\bar{\scriptO}_{i}$ denote the unobservable subspace of the pair $(C_{i},A)$ and let $P_{i} \in \doubleR^{n\times n}$ be the orthogonal projector onto $\bar{\scriptO}_{i}$. Define $n_{i}\defeq \dim \bar{\scriptO}_{i}$ and choose a matrix $V_{i}\in\doubleR^{n\times n_{i}}$ whose columns form an orthonormal basis of $\bar{\scriptO}_{i}$. Then, select a complementary matrix $M_{i}\in\doubleR^{n\times (n-n_{i})}$ so that the block matrix $[\,M_{i}\;\;V_{i}\,]$ is orthogonal. With this construction, note that $P_{i}=V_{i} V_{i}^{\top}$. Since $\bar{\scriptO}_{i}$ is $A$-invariant and contained in $\ker C_{i}$, we have $C_{i}V_{i}=0$ and $M_{i}^{\top} A V_{i}=0$. Thus,
    \vspace{-3pt}
    \begin{equation*}
        \begin{bmatrix}
            M_{i}^{\top} \\[3pt]
            V_{i}^{\top}
        \end{bmatrix}
        A
        \begin{bmatrix}
            M_{i} & V_{i}
        \end{bmatrix}
        =
        \begin{bmatrix}
            \bar{A}_{i} & 0 \\[3pt]
            * & *
        \end{bmatrix},
        \;\;
        C_{i} 
        \begin{bmatrix}
            M_{i} & V_{i}
        \end{bmatrix}
        =
        \begin{bmatrix}
            \bar{C}_{i} & 0
        \end{bmatrix},
        \vspace{-3pt}
    \end{equation*}
    where $\bar{A}_{i}\in\doubleR^{(n-n_{i})\times (n-n_{i})}$ and $\bar{C}_{i}\in\doubleR^{p_{i}\times (n-n_{i})}$. The reduced pair $(\bar C_{i},\bar A_{i})$ is observable \cite[Th. 6.O6]{Ch99}; hence, we can select a gain $\bar{L}_{i}\in\doubleR^{(n-n_{i})\times p_{i}}$ that makes $\bar{A}_{i} - \bar{L}_{i}\bar{C}_{i}$ Hurwitz with arbitrarily assigned eigenvalues. The observer gain is then defined by $L_{i} \defeq  M_{i} \bar{L}_{i}$. First, we impose the joint detectability of the system:
    \vspace{-3pt}
    \begin{asm1}\label{asm:jointobs}
        Defining $C\defeq[C_{1}^{\top},\ldots,C_{N}^{\top}]^{\top}$, the pair $(C,A)$ is detectable.
    \end{asm1}
    \vspace{-3pt}
    
    We now design a distributed observer on \eqref{eqn:discretized}, motivated by \cite{KLS25}. For each $i\in\scriptN$ and $k\in\doubleZ_{\geq0}$, let $\Delta_{i}^{k}\in\{0,1\}$ be a random variable, whose role will be specified later. Also, for each $k\in\doubleZ_{\geq0}$ and $i,j\in\scriptN$, let $w_{ij}^{k}\in\{0,1\}$ denote the random availability of the communication link from agent $j$ to agent $i$ at step $k$. We set $w_{ii}^{k} \equiv 0$ for all $i\in\scriptN$ and $k\in\doubleZ_{\geq0}$. The random graph induced by the link variables $\{w_{ij}^{k}\}$ at step $k$ is defined by $\scriptG[k] \defeq (\scriptN,\scriptE[k])$, with $\scriptE[k] \defeq \{(j,i)\in\scriptN\times\scriptN:w_{ij}^{k}=1\}$. The in-neighbor set of agent $i$ at step $k$ is defined as $\scriptN_{i}[k]\defeq \{j\in\scriptN:w_{ij}^{k}=1\}$.
    
    Let $(\Omega,\scriptF,\scriptP)$ be a probability space on which all random variables $\{\Delta_{i}^{k},w_{ij}^{k}:i,j\in\scriptN,\ k\in\doubleZ_{\geq0}\}$ are defined. Throughout this paper, the initial condition is assumed to be deterministic. For each $k \in \doubleN$, define the filtration
    \vspace{-3pt}
    \begin{equation*}
        \scriptF_{k} \defeq \sigma\left(\{ \Delta_{i}^{r},w_{ij}^{r}:i,j \in \scriptN,\;r = 0,\ldots, k-1\} \right),
        \vspace{-3pt}
    \end{equation*}
    with $\scriptF_{0} {\defeq} \{\emptyset,\Omega\}$. We impose the following condition on $\Delta_{i}^{k}$. 

    \begin{asm1}\label{asm:activation}
        For each $i \in \scriptN$, there exist finite constants $\ell_{i},\mathscr{L}_{i} \in \doubleR_{>0}$ with $\ell_{i} \leq \mathscr{L}_{i}$ such that for every $k \in \doubleZ_{\geq 0}$, 
        \vspace{-5pt}
        \begin{equation*}
            \frac{\ell_{i}}{\ell_{i}+1}
            \leq
            \scriptP(\Delta_{i}^{k}=0|\scriptF_{k})
            \leq
            \frac{\mathscr{L}_{i}}{\mathscr{L}_{i}+1}.
            \vspace{-5pt}
        \end{equation*}
    \end{asm1}

    We refer to such a $\Delta_{i}^{k}$ as an \emph{activation} random variable of agent $i$ at step $k$. 
    Before stating the assumptions on $w_{ij}^{k}$, we introduce a stochastic connectivity notion motivated by \cite[Def. 4]{SAJ15}: for $k_{1},k_{2} \in \doubleZ_{\geq 0}$ with $k_1\leq k_2$, denote the joint graph over the interval
    $[k_1,k_2]$ by $\scriptG([k_1,k_2])\defeq (\scriptN,\bigcup_{r=k_1}^{k_2}\scriptE[r])$.

    \begin{defn1}\label{defn:USSC}
        The random graph process $\{\scriptG[k]\}_{k \in \doubleZ_{\geq 0}}$ is said to be uniformly stochastically strongly connected (USSC) if there exist $B\in\doubleN$ and $q\in(0,1]$ such that
        \vspace{-3pt}
        \begin{equation*}
            \scriptP\left(
                \scriptG([mB,(m+1)B-1])
                \text{:strongly connected}\hspace{1pt}
                |
                \scriptF_{mB}
            \right)
            \geq q
            \vspace{-3pt}
        \end{equation*}
        holds almost surely, for every $m\in\doubleZ_{\geq0}$.
    \end{defn1}

    \vspace{-10pt}
    \begin{asm1}\label{asm:USSC}
        The process $\{\scriptG[k]\}_{k\in\doubleZ_{\geq0}}$ is USSC.
    \end{asm1}
    \vspace{-4pt}

    Assumption~\ref{asm:USSC} requires that, over each block $[mB,(m+1)B-1]$, the joint graph formed by all links appearing in that block is strongly connected with probability at least $q$, conditioned on the information available before the block starts. Thus, individual $\scriptG[k]$ may be sparse or disconnected, but the uniform lower bound $q>0$ on the conditional probability ensures that, over an infinite time horizon, infinitely many blocks $[mB,(m+1)B-1]$ have strongly connected joint graphs with probability one \cite[Th. 4.3.4]{Du19}. In contrast, the usual uniform joint strong connectivity condition requires the joint graph to be strongly connected over every block. Hence, USSC can be viewed as a natural stochastic relaxation of this condition, with the blockwise connectivity requirement being recovered, as a special case, almost surely when $q=1$. 

    Now, we introduce the distributed observer algorithm. At step $k$, agent $i$ updates its own \emph{local estimate} $\hat{x}_{i|i}[k]$ of the plant state using $y_{i}[k]$. Agent $i$ also stores buffer copies of the local estimates generated by other agents. Specifically, $\hat{x}_{i|j}[k]$ ($i \neq j$) denotes the \emph{buffer copy} stored by agent $i$ that originates from the local estimate of agent $j$. Such an estimate is initially created through communication from agent $j$ to agent $i$, and agent $i$ propagates it locally using the model dynamics until it is refreshed or removed.
    
    Specifically, for each agent $i$ and step $k$, let $\scriptM_{i}[k]$ denote the index set of agents available to agent $i$ at the beginning of step $k$. More precisely, $j \in \scriptM_{i}[k]$ if and only if $\hat{x}_{i|j}[k]$ has already been created and is available to agent $i$ by that time. From now on, we refer to $\scriptM_{i}[k]$ as the \emph{buffer-index set}. The buffer-index set is initialized as $\scriptM_{i}[0]=\{i\}$ and is updated at each subsequent step according to the rule specified later. At step $k$, agent $i$ forms the intermediate buffer-index set $\scriptM_{i}^{+}[k]\defeq\scriptM_{i}[k]\cup\scriptN_{i}[k]$. For each $i \in \scriptN$ and corresponding $j \in \scriptM_{i}^{+}[k]\setminus \{i\}$, the intermediate updates of its local estimate and buffer copies are given by
    \begin{equation*}
        \begin{aligned}
            \hat{x}_{i|i}[k^{+}]
            &=
            \hat{x}_{i|i}[k]
            +
            \alpha
            \Bigl[
                A\hat{x}_{i|i}[k]
                +
                L_{i}(y_{i}[k]-C_{i}\hat{x}_{i|i}[k])
            \Bigr],
            \\
            \hat{x}_{i|j}[k^{+}]
            &=
            \begin{cases}
                \hat{x}_{j|j}[k^{+}],& j \in \scriptN_{i}[k]\\
                (I_{n} + \alpha A)\hat{x}_{i|j}[k],& j \in \scriptM_{i}^{+}[k]\setminus (\scriptN_{i}[k] {\cup} \{i\}).
            \end{cases}
        \end{aligned}
    \end{equation*}
    The update of $\hat{x}_{i|i}[k^{+}]$ describes the \emph{local propagation} of agent $i$'s own estimate using its local measurement, whereas the update of $\hat{x}_{i|j}[k^{+}]$ describes either a \emph{communication-refresh} update $(j\in\scriptN_{i}[k])$ or a local propagation $(j\in\scriptM_{i}^{+}[k]\setminus(\scriptN_{i}[k]\cup\{i\}))$. $k^{+}$ denotes the intermediate stage after the above local propagation and communication-refresh operations. The local estimate at step $k+1$ and any buffer copies retained for the next step are then obtained through the \emph{projection-averaging} and \emph{buffer-reset} stage as
    \begin{align*}
        \hat{x}_{i|i}[k+1]
        &=
        \begin{cases}
            (I_{n}-P_{i})\hat{x}_{i|i}[k^{+}]\\
            +
            \displaystyle
            \frac{1}{|\scriptM_{i}^{+}[k]|}
            P_{i}
            \hspace{-5pt}\sum_{j\in\scriptM_{i}^{+}[k]}\hspace{-5pt}
            \hat{x}_{i|j}[k^{+}],
            &\Delta_{i}^{k}=0,\\
            \hat{x}_{i|i}[k^{+}],
            &
            \Delta_{i}^{k} = 1,
        \end{cases}\\
        \hat{x}_{i|j}[k+1] &= \hat{x}_{i|j}[k^{+}],\quad j \in \scriptM_{i}[k+1] \setminus \{i\},
    \end{align*}
    where the buffer-index set for the next step is given by
    \vspace{-3pt}
    \begin{equation*}
        \scriptM_{i}[k+1]
        =
        \begin{cases}
            \{i\},
            &\Delta_{i}^{k}=0,\\
            \scriptM_{i}^{+}[k],
            &\Delta_{i}^{k}=1.
        \end{cases}
        \vspace{-3pt}
    \end{equation*}
    Thus, the event $\Delta_{i}^{k}=0$ means that agent $i$ performs the projection-averaging update and resets all buffer copies, whereas $\Delta_{i}^{k}=1$ means that it skips the projection-averaging update and retains the accumulated buffer copies.
     
    We now impose some technical assumptions on $\Delta_{i}^{k}$:

    \begin{asm1}\label{asm:activation_ind}
        For every $k \in \doubleZ_{\geq 0}$, conditionally on $\scriptF_{k}$, the random variables $\Delta_1^{k},\ldots,\Delta_N^{k}$ are mutually independent. Moreover, for every $m \in \doubleZ_{\geq 0}$, the collection $\{\Delta_{i}^{r}:i\in\scriptN,\;r = mB,\ldots, (m+1)B-1\}$ is conditionally independent of $\{w_{ij}^{r}:i,j\in\scriptN,\;r = mB,\ldots, (m+1)B-1\}$ given $\scriptF_{mB}$.
    \end{asm1}

    Intuitively, Assumption~\ref{asm:activation_ind} is a natural modeling assumption: given the past, knowing which agents activate within a block does not change the probabilities of communication link availability in that block, and vice versa.

    The main result is stated below.

    \vspace{-3pt}
    \begin{thm1}\label{thm:main}
        Assume that Assumptions \ref{asm:jointobs}--\ref{asm:activation_ind} hold.
        Then, there exists $\bar{\alpha} \in \doubleR_{>0}$ such that for any $\alpha\in(0,\bar\alpha)$, 
        \vspace{-3pt}
        \begin{equation*}
            \lim_{k\to\infty}\|\hat x_{i|i}[k]-x[k]\|=0, \quad\text{a.s.,}\quad \forall i \in \scriptN,
            \vspace{-3pt}
        \end{equation*}
        with convergence rate $\gamma_{c} \in (0,1)$, i.e.,
        \vspace{-3pt}
        \begin{equation}\label{eqn:convergencerate}
            \limsup_{k\to\infty}
            \Big(
                \max_{i\in\scriptN}
                \|\hat x_{i|i}[k]-x[k]\|
            \Big)^{1/k}
            \leq
            \gamma_{c},\quad \text{a.s.}
        \end{equation}
    \end{thm1}

    We emphasize that the proposed observer updates do not require knowledge of $B$, $q$, $\ell_{i}$, or $\scriptL_{i}$ characterizing the random communication and activation processes.

    The generality of Theorem~\ref{thm:main} relies crucially on the freedom to choose the step size $\alpha$ sufficiently small. With the per-step communication and activation mechanisms kept fixed, reducing $\alpha$ provides more communication and projection-averaging opportunities per unit time on the continuous-time scale $t = k\alpha$. Indeed, the plant then evolves less between successive communication opportunities. Thus, the small step size requirement can be interpreted as requiring sufficiently frequent information exchange relative to plant evolution. We return to this interpretation in the subsequent analysis.

\section{Proof of Theorem~\ref{thm:main}}\label{sec:prf}


    Define $Q_{i} \defeq I_{n}-P_{i}$ and  $\Phi_{\alpha} \defeq I_{n}+\alpha A$. For $j\in\scriptM_{i}[k]$, define $e_{i|j}[k] \defeq \hat x_{i|j}[k]-x[k]$. Also, for $j \in \scriptM_{i}^{+}[k]$, define $e_{i|j}[k^{+}] \defeq \hat{x}_{i|j}[k^{+}] - x[k+1]$. Then, for each $i\in \scriptN$ and each $j \in \scriptM_{i}^{+}[k]\setminus\{i\}$, the intermediate local propagation and communication-refresh updates are given by
    \begin{equation*}
        \begin{aligned}
            e_{i|i}[k^{+}] &= \bigl(I_{n}+\alpha(A-L_{i}C_{i})\bigr)e_{i|i}[k],\\
            e_{i|j}[k^{+}]
            &=
            \begin{cases}
                e_{j|j}[k^{+}],
                & j\in\scriptN_{i}[k],
                \\
                \Phi_{\alpha} e_{i|j}[k],
                & j\in\scriptM_{i}^{+}[k]\setminus(\scriptN_{i}[k]\cup\{i\}).
            \end{cases}
        \end{aligned}
    \end{equation*}
    The projection-averaging and buffer-reset stage is given by
    \begin{equation*}
        \begin{aligned}
            e_{i|i}[k+1]
            &=
            \begin{cases}
                Q_{i} e_{i|i}[k^{+}]\\
                +
                \displaystyle
                \frac{1}{|\scriptM_{i}^{+}[k]|}
                P_{i}\hspace{-5pt}
                \sum_{j\in\scriptM_{i}^{+}[k]}
                \hspace{-5pt}e_{i|j}[k^{+}],
                &
                \Delta_{i}^{k}=0,
                \\
                e_{i|i}[k^{+}],
                &
                \Delta_{i}^{k} = 1,
            \end{cases}\\
            e_{i|j}[k+1]&=e_{i|j}[k^{+}],\quad j\in\scriptM_{i}[k+1]\setminus\{i\}.
        \end{aligned}
    \end{equation*}
    For convenience, write $e_{i}[k] \defeq e_{i|i}[k]$ and $e_{i}[k^{+}] \defeq e_{i|i}[k^{+}]$. Also, we refer to $e_{i}[k]$ as the local estimation error, and to $e_{i|j}[k]$ with $j \in \scriptM_{i}[k] \setminus \{i\}$ as a buffer error. 
    
    \subsection{Error Dynamics: Locally Observable Part}
        
        For $j\in\scriptM_{i}[k]\setminus\{i\}$, define the last refresh time of the copy originating from agent $j$ stored by agent $i$ as
        \vspace{-3pt}
        \begin{equation*}
            \tau_{i|j}[k]
            \defeq
            \max
            \left\{
                \tau:\hspace{-4pt}
                \begin{array}{l}
                    \tau\in\{0,\ldots,k-1\},\ j\in\scriptN_{i}[\tau],\\[1mm]
                    j\in\scriptM_{i}[s]\ \text{for all }s=\tau+1,\ldots,k
                \end{array}
            \right\}.
            \vspace{-3pt}
        \end{equation*}
        Since the buffer error is propagated locally during steps $\tau_{i|j}[k]+1,\ldots,k-1$, we obtain that 
        \vspace{-4pt}
        \begin{equation}\label{eqn:decomposition_detailed}
            e_{i|j}[k]
            =
            \Phi_{\alpha}^{k-\tau_{i|j}[k]-1}(P_{j} + Q_{j})e_{j}[\tau_{i|j}[k]^{+}].
            \vspace{-4pt}
        \end{equation}
        Then, $e_{i|j}[k]$ can be decomposed into sum of a locally unobservable component $u_{i|j}[k]$ and a residual $r_{i|j}[k]$:
        \vspace{-4pt}
        \begin{equation}\label{eqn:decomposition}
            \begin{aligned}
                u_{i|j}[k] &\defeq \Phi_{\alpha}^{k-\tau_{i|j}[k]-1}P_{j} e_{j}[\tau_{i|j}[k]^{+}],\\
                r_{i|j}[k] &\defeq \Phi_{\alpha}^{k-\tau_{i|j}[k]-1}Q_{j} e_{j}[\tau_{i|j}[k]^{+}].
            \end{aligned}
            \vspace{-4pt}
        \end{equation}
        Since $\bar{\scriptO}_{j}$ is $A$-invariant, $u_{i|j}[k]\in\bar{\scriptO}_{j}$. Moreover, since $Q_{j}P_{j} = 0$, we obtain $Q_{j}e_{j}[\tau_{i|j}[k]^{+}] = Q_{j}e_{j}[\tau_{i|j}[k]+1]$. We now claim that $Q_{i}e_{i}[k]$, which is the locally observable component of the local estimation error $e_{i}[k]$, and the residual $r_{i|j}[k]$ converge exponentially to zero. 
        
        \begin{lem1}\label{lem:local_fast_part}
            There exists a deterministic $\alpha_{1} \in \doubleR_{>0}$ such that, for every $\alpha\in(0,\alpha_1)$, the following statements hold.\footnote{Note that the maximum over an empty set is understood as zero.}
            \begin{enumerate}[leftmargin = 12pt]
                \item There exist deterministic constants $c_Q \in \doubleR_{>0}$ and $\gamma_Q\in(0,1)$ such that
                    \vspace{-4pt}
                    \begin{equation*}
                        \max_{i\in\scriptN}\|Q_{i} e_{i}[k]\|\leq c_Q\gamma_Q^{k} \max_{i\in\scriptN}\|Q_{i} e_{i}[0]\|,\quad \forall k \in \doubleZ_{\geq 0}.
                        \vspace{-4pt}
                    \end{equation*}
                    Note that the constant $\gamma_{Q}$ can be chosen to satisfy
                    \vspace{-3pt}
                    \begin{equation}\label{eqn:gamma_Q}
                        \gamma_Q\in\bigg(\max_{\substack{i\in\scriptN\\\lambda\in\frakturS(\bar A_{i}-\bar L_{i}\bar C_{i})}}|1+\alpha\lambda|,\,1\bigg).
                        \vspace{-3pt}
                    \end{equation}
                    
                \item There exists $\Omega_{1} \subseteq \Omega$ with $\scriptP(\Omega_{1}) = 1$, such that for each $\omega \in \Omega_{1}$ and $\gamma_{R} \in (\gamma_{Q},1)$, $r_{i|j}[k]$ satisfies 
                    \vspace{-3pt}
                    \begin{equation*}
                        \max_{\substack{i\in\scriptN,j\in\scriptM_{i}[k]\setminus \{i\}}}
                        \|r_{i|j}[k]\|
                        \leq
                        c_R(\omega)\gamma_R^{k},\quad \forall k \in \doubleZ_{\geq 0},
                        \vspace{-3pt}
                    \end{equation*}
                    with $c_R(\omega,\gamma_{R}) \in \doubleR_{>0}$ finite. 
            \end{enumerate}    
        \end{lem1}

        \begin{proof}
            Refer to Appendix~\ref{app:local_fastpart}.
        \end{proof}

        We note that the dependence of $c_{R}(\omega)$ on $\gamma_{R}$ and the dependence of $c_{Q},\gamma_{Q},\Omega_{1},c_{R}(\omega),\gamma_{R}$ on $\alpha$ are suppressed for notational simplicity.

    \subsection{Error Dynamics: Locally Unobservable Part}

    We now isolate the unobservable components remaining after removing the exponentially decaying locally observable components and residuals identified in Lemma~\ref{lem:local_fast_part}. The resulting homogeneous reduced dynamics is described by 
    \begin{equation*}
        \begin{aligned}
            \bar u_{i|i}[k^{+}]
            &=
            \Phi_{\alpha} \bar u_{i|i}[k],
            \\
            \bar u_{i|j}[k^{+}]
            &=
            \begin{cases}
                \bar u_{j|j}[k^{+}], & j\in\scriptN_{i}[k], \\
                \Phi_{\alpha} \bar u_{i|j}[k], & j\in\scriptM_{i}^{+}[k]\setminus(\scriptN_{i}[k]\cup\{i\}),
            \end{cases}\\
            \bar u_{i|i}[k+1] &=
            \begin{cases}
                \displaystyle
                \frac{1}{|\scriptM_{i}^{+}[k]|}
                P_{i}\hspace{-5pt}
                \sum_{j\in\scriptM_{i}^{+}[k]}
                \hspace{-5pt}
                \bar u_{i|j}[k^{+}],
                &
                \Delta_{i}^{k}=0,
                \\
                \bar u_{i|i}[k^{+}],
                &
                \Delta_{i}^{k} = 1,
            \end{cases}\\
            \bar u_{i|j}[k+1] &= \bar u_{i|j}[k^{+}],\quad j\in\scriptM_{i}[k+1]\setminus\{i\},
        \end{aligned}
    \end{equation*}
    where $\bar{u}_{i|i}[0] \in \bar{\scriptO}_{i}$. Note that initializing the $\bar{u}_{i|j}$ components is unnecessary, because $\scriptM_{i}[0]\setminus \{i\} = \emptyset$. Since each subspace $\bar{\scriptO}_{i}$ is $A$-invariant and $P_{i}$ maps onto $\bar{\scriptO}_{i}$, the above dynamics preserves $\bar{u}_{i|i}[k]\in\bar{\scriptO}_{i}$ for every $i\in\scriptN$ and $\bar{u}_{i|j}[k]\in\bar{\scriptO}_{j}$ for every $j\in\scriptM_{i}[k]\setminus\{i\}$. For convenience, write $\bar{u}_{i}[k] \defeq \bar{u}_{i|i}[k]$ and $\bar{u}_{i}[k^{+}] \defeq \bar{u}_{i|i}[k^{+}]$. Let $\bar\xi[k]$ be the stacked vector 
    \vspace{-4pt}
    \begin{equation*}
        \bar\xi[k] \defeq \operatorname{col} \left( \bar u_1[k],\ldots,\bar u_N[k], \{\bar u_{i|j}[k]\}_{i,j\in\scriptN,\ j\neq i} \right) \in\doubleR^{N^{2}n},
        \vspace{-4pt}
    \end{equation*} 
    where $\bar{u}_{i|j}[k]$ with $j \notin \scriptM_{i}[k]$ is set to zero and the ordered collection is taken in a fixed lexicographic order. For each realization $\omega \in \Omega$, let $\scriptT_{k}^{\alpha}(\omega)$ denote the one-step linear transition of the homogeneous reduced dynamics, given as
    \vspace{-5pt}
    \begin{equation}\label{eqn:reduced_dynamics}
        \bar\xi[k+1]=\scriptT_{k}^{\alpha}(\omega)\bar\xi[k].
        \vspace{-5pt}
    \end{equation} 
    The remaining proof of Theorem~\ref{thm:main} proceeds by expressing the locally unobservable error dynamics as \eqref{eqn:reduced_dynamics} with exponentially decaying additive perturbations, and conducting the convergence analysis. Before proceeding to the specific step, given $\omega \in \Omega$ and a stacked vector
    \vspace{-5pt} 
    \begin{equation}\label{eqn:stack}
        v = \operatorname{col}\left(v_1,\ldots,v_N,\{v_{i|j}\}_{i,j\in\scriptN,\ j\neq i}\right)
        \vspace{-5pt}
    \end{equation}
    with the same lexicographic order as in $\bar\xi[k]$, we say $v$ is $(k,\omega)$-structurally compatible if $v_{i}\in\bar{\scriptO}_{i}$ for every $i\in\scriptN$, $v_{i|j}\in\bar{\scriptO}_{j}$ whenever $j\in\scriptM_{i}[k]\setminus\{i\}$, and $v_{i|j}=0$ whenever $j \notin \scriptM_{i}[k]$. Both $\scriptT_{k}^{\alpha}(\omega)$ and the notion of structural compatibility depend on $\omega$. For notational simplicity, however, we suppress this dependence on $\omega$ if no ambiguity arises.

    Define $u_{i}[k]\defeq P_{i} e_{i}[k]$. Also, for $j \in \scriptM_{i}[k] \setminus\{i\}$, write $e_{i|j}[k]=u_{i|j}[k]+r_{i|j}[k]$ as in \eqref{eqn:decomposition}; for $j \notin \scriptM_{i}[k]$, set $u_{i|j}[k]=0$. Using the stacking convention of $\bar\xi[k]$, define
    \vspace{-6pt}
    \begin{equation*}
        \xi[k]
        \defeq
        \operatorname{col}
        \left(
            u_1[k],\ldots,u_N[k],
            \{u_{i|j}[k]\}_{i,j\in\scriptN,\ j\neq i}
        \right)
        \in\doubleR^{N^{2}n},
        \vspace{-3pt}
    \end{equation*} 
    where the ordered collection is taken in a fixed lexicographic order. Note that $\xi[k]$ consists of unobservable components of $e_{i}[k]$ and $e_{i|j}[k]$, and is $k$-structurally compatible. We first express the dynamics of $\xi[k]$ using $\scriptT_{k}^{\alpha}$ defined in \eqref{eqn:reduced_dynamics}:

    \begin{lem1}\label{lem:unobservabledynamics}
        $\xi[k]$ follows the one-step transition $\scriptT_{k}^{\alpha}$, with additive perturbation terms $d[k]$ given as
        \vspace{-3pt}
        \begin{equation}\label{eqn:robust}
            \xi[k+1]=\scriptT_{k}^{\alpha}\xi[k]+d[k],
            \vspace{-3pt}
        \end{equation}
        where $d[k]$ is induced by $Q_{i}e_{i}[k]$, $Q_{i}e_{i}[k^{+}]$, and $r_{i|j}[k]$, and is $(k+1)$-structurally compatible.
    \end{lem1}

    \begin{proof}
        Refer to Appendix~\ref{app:unobservabledynamics}.
    \end{proof}
    \vspace{-8pt}
    
    The projection-averaging update gives $Q_{i} e_{i}[k^{+}]=Q_{i} e_{i}[k+1]$. Hence, the terms $Q_{i} e_{i}[k^{+}]$ are also covered by the exponential estimate in Lemma~\ref{lem:local_fast_part}. Applying the above observation, with the form of $d[k]$ in Appendix~\ref{app:unobservabledynamics}, for the same $\alpha$ and $\gamma_{R}$ in Lemma~\ref{lem:local_fast_part}, there exists a random variable $c_{d}(\omega)$, finite on $\Omega_{1}$, such that
    \vspace{-4pt}
    \begin{equation}\label{eqn:gamma_zeta}
        \|d[k]\|\leq c_{d}(\omega)\gamma_{R}^{k},\quad \forall k \geq 0.   
        \vspace{-4pt}
    \end{equation}
    As in Lemma~\ref{lem:local_fast_part}, the dependence of $c_{d}(\omega)$ on $\alpha$ and $\gamma_{R}$ is suppressed for notational simplicity.

    We now analyze the stability of \eqref{eqn:robust}. Let $\scriptC_{m} \subseteq \Omega$ be the event that the joint graph $\scriptG([mB,(m+1)B-1])$ is strongly connected, and $\scriptD_{m} \subseteq \Omega$ be the event that $\Delta_{i}^{r}=1$ for all $i\in\scriptN$ and $r=mB,\ldots,(m+1)B-2$, while $\Delta_{i}^{(m+1)B-1}=0$ for all $i\in\scriptN$. Set the event $\scriptH_{m}\defeq\scriptC_{m}\cap\scriptD_{m}$. We first show the contraction property of $\scriptT_{k}^{\alpha}$ under the event $\scriptK_{m} \defeq \bigcap_{s=mN}^{(m+1)N-1}\scriptH_{s}$. 
    For convenience, define $\scriptS\defeq \bigcap_{i=1}^N\bar{\scriptO}_{i}$. Let $\Pi_{\scriptS}$ denote the orthogonal projector onto $\scriptS$, and define two seminorms
    \begin{equation*}
        \begin{aligned}
            &\|z\|_{\scriptS}\defeq \|(I_{n}-\Pi_{\scriptS})z\|,\\
            &\|\operatorname{col}(z_{1},\ldots,z_{N})\|_{\scriptS,\max} \defeq \max_{i \in \{1,\ldots,N\}}\|z_{i}\|_{\scriptS},
        \end{aligned}
        \vspace{-4pt}
    \end{equation*}
    where $z,z_{1},\ldots,z_{N} \in \doubleR^{n}$. 
    \begin{lem1}\label{lem:homogeneous}
        There exist deterministic $\alpha_{b} \in \doubleR_{>0}$ and $\rho_{b} \in (0,1)$ such that for every $m\in\doubleZ_{\geq0}$, every $\omega\in\scriptK_{m}$, and every $\alpha\in(0,\alpha_{b})$,
        \vspace{-6pt}
        \begin{equation}\label{eqn:alphacontraction}
            \|
                \frakturB_{(m+1)N-1}^{\alpha}
                \cdots
                \frakturB_{mN+1}^{\alpha}\mathbf{z}
            \|_{\scriptS,\max}
            \leq
            \rho_{b}\|\mathbf{z}\|_{\scriptS,\max},
            \vspace{-4pt}
        \end{equation}
        where $\mathbf{z}=\operatorname{col}(z_{1},\ldots,z_{N})$ with $z_{i}\in\bar{\scriptO}_{i}$, and $\frakturB_{s}^{\alpha}$ is the mapping given by $\frakturB_{s}^{\alpha} \mathbf{z}\defeq (\scriptT_{(s+1)B-1}^{\alpha}\cdots\scriptT_{sB}^{\alpha}\xi_{\mathbf{z}})_{1:Nn}$, where $\xi_{\mathbf{z}}\defeq\operatorname{col}\left(\mathbf{z},0\right)\in\doubleR^{N^{2}n}$ and $(\cdot)_{1:Nn}$ denotes the truncation that retains the first $Nn$ coordinates. 
    \end{lem1}

    \begin{proof}
        Refer to Appendix~\ref{prf:lem:homogeneous}.
    \end{proof}

    \vspace{-5pt}

    We note that this lemma contains the key idea for removing the commutativity condition $AP_{i}=P_{i}A$ required in \cite{KLS25}; this point will be discussed in more detail later.

    From $\xi[k]$, define
    \vspace{-4pt}
    \begin{equation*}
        \Xi[k]
        \defeq
        \max
        \left\{
            \|u_{i}[k]\|_{\scriptS},
            \|u_{i|j}[k]\|_{\scriptS}
            :
            i{\in}\scriptN,j{\in}\scriptM_{i}[k]\setminus\{i\}
        \right\}.
        \vspace{-4pt}
    \end{equation*}
    Apart from the propagation by $\Phi_{\alpha}$, the operations in $\scriptT_{k}^{\alpha}$ do not increase the value of $\Xi[k]$. Indeed, communication-refresh only duplicates an existing component, and buffer-reset only removes buffer components. Moreover, the projection-averaging step is nonexpansive in $\|\cdot\|_{\scriptS}$.     
    Since $\scriptS$ is $A$-invariant, $(I_{n}-\Pi_{\scriptS})A\Pi_{\scriptS}=0$. Define 
    \vspace{-4pt}
    \begin{equation}\label{eqn:c_A}
        c_A\defeq\|(I_{n}-\Pi_{\scriptS})A(I_{n}-\Pi_{\scriptS})\| \in \doubleR_{\geq 0}.    
        \vspace{-4pt}
    \end{equation} 
    Then, for every $\alpha\in \doubleR_{\geq 0}$,
    \vspace{-4pt}
    \begin{equation*}
        \begin{aligned}
            \|\Phi_{\alpha} z\|_{\scriptS}
            &=
            \|(I_{n}-\Pi_{\scriptS})(I_{n}+\alpha A)z\|
            \\
            &=
            \|(I_{n}-\Pi_{\scriptS})z
            +
            \alpha(I_{n}-\Pi_{\scriptS})A(I_{n}-\Pi_{\scriptS})z\|
            \\
            &\leq
            (1+c_A\alpha)\|z\|_{\scriptS}.
        \end{aligned}
        \vspace{-4pt}
    \end{equation*}
    Therefore, there exists a constant $c_{\Xi} \in \doubleR_{>0}$, independent of $k$ and $T$, such that
    \vspace{-4pt}
    \begin{equation}\label{eqn:T_estimate}
        \begin{aligned}
            \Xi[k+T]
            &\leq
            (1+c_A\alpha)^T \Xi[k]\\
            &\quad +
            c_{\Xi}
            \sum_{r=k}^{k+T-1}
            (1+c_A\alpha)^{k+T-1-r}\|d[r]\|.
        \end{aligned}
        \vspace{-4pt}
    \end{equation}
    With \eqref{eqn:alphacontraction} of Lemma~\ref{lem:homogeneous} and adding the contribution of $d[k]$ through \eqref{eqn:T_estimate}, we obtain
    \vspace{-4pt}
    \begin{equation}\label{eqn:scalar_recursion}
        \Xi[(m+1)NB]
        \leq
        a_{\alpha} \rho_{b}^{\mathbf{1}_{\scriptK_{m}}} \Xi[mNB]
        +
        \beta_{m},
        \vspace{-4pt}
    \end{equation}
    where $a_{\alpha}\defeq(1+c_A\alpha)^{NB}\in \doubleR_{>0}$ and $\beta_{m}$ is given by
    \vspace{-6pt}
    \begin{equation*}
        \beta_{m}
        \defeq
        c_{\Xi}
        \sum_{k=mNB}^{(m+1)NB-1}
        (1+c_A\alpha)^{(m+1)NB-1-k}\|d[k]\|.
        \vspace{-6pt}
    \end{equation*}

    We next estimate the frequency of the events $\scriptK_{m}$. By Assumptions~\ref{asm:activation}--\ref{asm:activation_ind},
    \vspace{-5pt}
    \begin{equation*}
        \scriptP(\scriptH_{m}|\scriptF_{mB})
        \geq
        q
        \product{i=1}{N}{\frac{\ell_{i}}{\ell_{i} + 1}\bigg(\frac{1}{\mathscr{L}_{i}+1}\bigg)^{B-1}}
        \eqdef p_H\in \doubleR_{>0}.
        \vspace{-5pt}
    \end{equation*}
    Then, 
    \vspace{-4pt}
    \begin{equation}\label{eqn:pk}
        \scriptP(\scriptK_{m}|\scriptF_{mNB})\geq p_H^N\eqdef p_K\in \doubleR_{>0}.  
        \vspace{-4pt}
    \end{equation}
    Given $r \in \doubleZ_{\geq 0}$, set $M_{\nu}^{r} \defeq \sum_{s=r}^{r+\nu-1}(\mathbf{1}_{\scriptK_{s}} - \doubleE[\mathbf{1}_{\scriptK_{s}}|\scriptF_{sNB}])$. It is clear that $M_{\nu}^{r}$ is a martingale with respect to the filtration $\{\scriptF_{(r+\nu)NB}\}_{\nu \in \doubleZ_{\geq 0}}$. Since $\doubleE[\mathbf{1}_{\scriptK_{m}}|\scriptF_{mNB}]\geq p_K$,
    \vspace{-4pt}
    \begin{equation*}
        \bigg\{\sum_{s=r}^{r+\nu-1}\mathbf{1}_{\scriptK_{s}}\leq\frac{p_K}{2}\nu\bigg\} \subseteq \bigg\{M_{\nu}^{r} \leq -\frac{p_{K}}{2}\nu \bigg\}.
        \vspace{-4pt}
    \end{equation*}
    Applying Azuma's inequality \cite[Th. D.7]{MRT18} gives
    \vspace{-4pt}
    \begin{equation*}
        \scriptP
        \bigg(
            \sum_{s=r}^{r+\nu-1}\mathbf{1}_{\scriptK_{s}}
            \leq
            \frac{p_K}{2}\nu
        \bigg)
        \leq
        \scriptP
        \bigg(
            M_{\nu}^{r} \leq -\frac{p_{K}}{2}\nu
        \bigg)
        \leq
        e^{-c_K\nu},
        \vspace{-4pt}
    \end{equation*}   
    where $c_K \defeq p_{K}^{2}/8 \in \doubleR_{>0}$. Choose a deterministic constant $c_{0} \in \doubleR_{>0}$ such that $c_{0}>3/c_K$. For each $M\in\doubleN$, define 
    \vspace{-4pt}
    \begin{align*}
        \scriptB_M
        \defeq
        \Bigg\{
        \begin{aligned}
            &\exists\, 0 \leq r < m \leq M
            \text{ s.t. }
            m-r \ge c_{0} \log(M+2), \\[-2pt]
            &\textstyle\sum_{s=r}^{m-1} \mathbf{1}_{\scriptK_{s}}
            <
            p_K(m-r)/2
        \end{aligned}
        \Bigg\}.
        \vspace{-4pt}
    \end{align*}
    By the preceding Azuma estimate, 
    \vspace{-3pt}
    \begin{equation*}
        \begin{aligned}
            \scriptP(\scriptB_M)
            &\leq
            (M+1)^{2}
            \exp\bigl(-c_Kc_{0}\log(M+2)\bigr)
            \\
            &=
            (M+1)^{2}(M+2)^{-c_Kc_{0}}.
        \end{aligned}
        \vspace{-3pt}
    \end{equation*}
    Since $c_Kc_{0}>3$, the right-hand side is summable in $M$. Therefore, by the Borel--Cantelli lemma \cite[Th. 2.3.1]{Du19}, there exists an event $\Omega_2$ with $\scriptP(\Omega_2)=1$ such that, for every $\omega\in\Omega_2$, there exists $M_2(\omega) \in \doubleN$ for which the event $\scriptB_M$ does not occur for all $M\geq M_2(\omega)$. Taking $M=m$, we obtain that for every $\omega\in\Omega_2$, every $m\geq M_2(\omega)$, and every $r\in\{0,\ldots,m-1\}$ satisfying
    $m-r\geq c_{0}\log(m+2)$,
    \vspace{-4pt}
    \begin{equation*}
        \sum_{s=r}^{m-1}\mathbf{1}_{\scriptK_{s}}
        \geq
        \frac{p_K}{2}(m-r),
        \vspace{-4pt}
    \end{equation*}
    i.e., almost surely, the events $\scriptK_{s}$ occur with frequency at least $p_{K}/2$ over every interval $[r,m)$ satisfying $m-r \geq c_{0}\log(m+2)$, for all sufficiently large $m$.

    Turning back to \eqref{eqn:scalar_recursion}, define
    \vspace{-4pt}
    \begin{equation}\label{eqn:widealpha}
        \widehat{\alpha}
        \defeq
        \frac{\rho_{b}^{-p_{K}/(2NB)}-1}{c_{A}}
        \in \doubleR_{>0}\cup\{+\infty\},
        \vspace{-4pt}
    \end{equation}
    with the convention that $\widehat\alpha=+\infty$ when $c_A=0$. Fix any $\alpha \in \doubleR_{>0}$ satisfying
    \vspace{-4pt}
    \begin{equation}\label{eqn:alpha}
        \alpha\in(0,\min\{\alpha_{1},\alpha_{b},\widehat{\alpha}\}).
        \vspace{-4pt}
    \end{equation}
    In what follows, constants may depend on this fixed $\alpha$, and this dependence is suppressed unless displayed explicitly. Since $a_{\alpha}\rho_{b}^{p_{K}/2}<1$, define the deterministic constants $\lambda\defeq-\log(a_{\alpha}\rho_{b}^{p_{K}/2})\in\doubleR_{>0}$ and $\kappa\defeq c_{0}(\log a_{\alpha}+\lambda)\in\doubleR_{>0}$. Also, choose a deterministic constant $\bar{\gamma}_{R}\in(\gamma_{R}^{NB},1)$. 
    
    Now fix any $\omega\in\Omega_{1}\cap\Omega_{2}$. Since $\omega\in\Omega_{1}$, \eqref{eqn:gamma_zeta} gives $\|d[k]\|\leq c_{d}(\omega)\gamma_{R}^{k}$ for all $k\in\doubleZ_{\geq0}$. By the definition of $\beta_{m}$, there exists a finite constant $c_{\beta}(\omega)\in\doubleR_{>0}$ such that $\beta_{m}\leq c_{\beta}(\omega)\bar{\gamma}_{R}^{m}$ for all $m\in\doubleZ_{\geq0}$.

    We now claim that there exists $c_P(\omega) \in \doubleR_{>0}$ such that
    \vspace{-6pt}
    \begin{equation}\label{eqn:probabilistic_bound}
        \prod_{s=r}^{m-1}
        a_{\alpha}\rho_{b}^{\mathbf{1}_{\scriptK_{s}}}
        \leq
        c_P(\omega)(m+2)^{\kappa} e^{-\lambda(m-r)}.
        \vspace{-6pt}
    \end{equation}
    Indeed, if $m\geq M_2(\omega)$ and
    $m-r\geq c_0\log(m+2)$, then
    $\sum_{s=r}^{m-1}\mathbf{1}_{\scriptK_{s}}\geq(p_K/2)(m-r)$, and hence the product is bounded by $e^{-\lambda (m-r)}$. If $m-r < c_0\log(m+2)$, then
    \vspace{-6pt}
    \begin{equation*}
        \prod_{s=r}^{m-1}
        a_{\alpha}\rho_{b}^{\mathbf{1}_{\scriptK_{s}}}
        \leq
        a_{\alpha}^{m-r}
        \leq
        (m+2)^{\kappa} e^{-\lambda (m-r)}.
        \vspace{-6pt}
    \end{equation*}
    Finally, the finitely many pairs $(r,m)$ with
    $m<M_2(\omega)$ are absorbed into $c_P(\omega)$, proving \eqref{eqn:probabilistic_bound}. Iterating \eqref{eqn:scalar_recursion} gives
    \vspace{-6pt}
    \begin{equation}\label{eqn:scalarrecursion}
        \begin{aligned}
            &\Xi[mNB]
            \leq
            c_P(\omega)(m+2)^{\kappa} e^{-\lambda m}\Xi[0]
            \\[-2pt]
            &\quad+
            c_P(\omega)c_{\beta}(\omega)(m+2)^{\kappa}
            \sum_{r=0}^{m-1}
            e^{-\lambda(m-1-r)}\bar\gamma_{R}^{r}.
        \end{aligned}
        \vspace{-6pt}
    \end{equation}
    For the right-hand side of \eqref{eqn:scalarrecursion} converges exponentially to zero, $ \Xi[mNB] \to 0$ almost surely. The exponential convergence at intermediate times follows from \eqref{eqn:T_estimate} with $T\leq NB$, because $\Xi[mNB]\to0$ and $\|d[k]\|$ decays exponentially.    

    It remains to control the component in $\scriptS$, for the convergence of $\Xi[k]$ only implies the convergence toward the set $\scriptS$. Suppose that $\scriptS\neq\{0\}$ (the case when $\scriptS = \{0\}$ is trivial). Since $\scriptS$ is $A$-invariant, the restriction $A|_{\scriptS}:\scriptS\to\scriptS$ is well-defined. Furthermore, Assumption~\ref{asm:jointobs} implies that $A|_{\scriptS}$ is Hurwitz. 
    Thus, there exists a self-adjoint positive-definite linear operator $H_{\scriptS}:\scriptS\to\scriptS$ satisfying
    \vspace{-6pt}
    \begin{equation}\label{eqn:lyapunov}
        (A|_{\scriptS})^{*}H_{\scriptS} + H_{\scriptS}(A|_{\scriptS}) = -\operatorname{id}_{\scriptS},
        \vspace{-6pt}
    \end{equation}
    where $(A|_{\scriptS})^{*}$ denotes the adjoint of $A|_{\scriptS}$ with respect to the Euclidean inner product restricted to $\scriptS$, and $\mathrm{id}_{\scriptS}$ denotes the identity operator on $\scriptS$. Define the norm $\overline{\|s\|}_{\scriptS}\defeq\sqrt{s^{\top}H_{\scriptS}s}$ for $s\in\scriptS$. Similarly to $\Xi[k]$, define
    \vspace{-3pt}
    \begin{equation*}
        \begin{aligned}
            &Y[k]\\
            &{\defeq}
            \max
            \{
                \overline{\|\Pi_{\scriptS}u_{i}[k]\|}_{\scriptS},
                \overline{\|\Pi_{\scriptS}u_{i|j}[k]\|}_{\scriptS}
                :
                i{\in}\scriptN,j{\in}\scriptM_{i}[k]\setminus\{i\}
            \}.
        \end{aligned}
        \vspace{-3pt}
    \end{equation*}
    Since $\scriptS$ is $A$-invariant, for every $s\in\scriptS$,
    \vspace{-3pt}
    \begin{equation*}
        \begin{aligned}
            &\overline{\|\Phi_{\alpha} s\|}_{\scriptS}^{\,2}
            = s^{\top}(I+\alpha A|_{\scriptS})^{\top}H_{\scriptS}(I+\alpha A|_{\scriptS})s
            \\
            &= \overline{\|s\|}_{\scriptS}^{\,2} - \alpha\|s\|^{2} + \alpha^{2}s^{\top}
            (A|_{\scriptS})^{\top}H_{\scriptS}(A|_{\scriptS})s
            \\
            &\leq\Big[1 - \frac{\alpha}{\frakturS_{\max}(H_{\scriptS})} + \alpha^{2} \left\|H_{\scriptS}^{1/2}(A|_{\scriptS})H_{\scriptS}^{-1/2}\right\|^{2}
            \Big]
            \overline{\|s\|}_{\scriptS}^{\,2},
        \end{aligned}
        \vspace{-3pt}
    \end{equation*}
    where $\frakturS_{\max}(H_{\scriptS})$ denotes the largest eigenvalue of $H_{\scriptS}$. 
    
    Defining $\mu \defeq 1/(4\frakturS_{\max}(H_{\scriptS}))$, set
    \vspace{-4pt}
    \begin{equation}\label{eqn:alpha_s}
        \alpha_{s} \defeq 2\mu\|H_{\scriptS}^{1/2}(A|_{\scriptS})H_{\scriptS}^{-1/2}\|^{-2}.
        \vspace{-4pt}
    \end{equation}
    Then, for all $\alpha \in (0,\alpha_{s})$,
    \vspace{-4pt}
    \begin{equation*}
        \begin{aligned}
            \overline{\|\Phi_{\alpha} s\|}_{\scriptS}^{2}
            \leq (1 - 2\mu\alpha)\overline{\|s\|}_{\scriptS}^{2}
            \leq \left(1 - \mu\alpha\right)^{2}\overline{\|s\|}_{\scriptS}^{2},
            \quad
            \forall s\in\scriptS.
        \end{aligned}
        \vspace{-4pt}
    \end{equation*}
    From \eqref{eqn:lyapunov}, it can be directly obtained that $\alpha_{s} < \mu^{-1}$, i.e., $1- \mu\alpha \in (0,1)$ when $\alpha \in (0,\alpha_{s})$. 
    
    Given any $z\in\doubleR^{n}$, 
    \vspace{-4pt}
    \begin{equation*}
        \begin{aligned}
            \Pi_{\scriptS}\Phi_{\alpha} z
            &=
            \Phi_{\alpha}\Pi_{\scriptS}z
            +
            \alpha\Pi_{\scriptS}A(I_{n}-\Pi_{\scriptS})z.
        \end{aligned}
        \vspace{-4pt}
    \end{equation*}
    Hence, for some constant $c_{q}\in \doubleR_{>0}$,
    \vspace{-4pt}
    \begin{equation*}
        \overline{\|\Pi_{\scriptS}\Phi_{\alpha} z\|}_{\scriptS}
        \leq
        (1-\mu\alpha)\overline{\|\Pi_{\scriptS}z\|}_{\scriptS}
        +
        \alpha c_{q}\|z\|_{\scriptS},\quad \forall \alpha \in (0,\alpha_{s}).
        \vspace{-4pt}
    \end{equation*}
    Since the remaining operations (e.g., communication-refresh, buffer-reset, and projection-averaging) do not increase the maximum of the $\scriptS$-components and the structurally compatible input $d[k]$ contributes at most a constant multiple of $\|d[k]\|$, there exist constants $c_{S},c_{d}'\in \doubleR_{>0}$ such that
    \vspace{-3pt}
    \begin{equation}\label{eqn:recursionforY}
        Y[k+1]\leq(1-\mu\alpha)Y[k] + c_{S} \Xi[k] + c_{d}'\|d[k]\|.
        \vspace{-3pt}
    \end{equation}
    If $\Xi[k],\|d[k]\|\to0$ exponentially, $Y[k]\to0$ exponentially. 

    From \eqref{eqn:alpha} and \eqref{eqn:alpha_s}, set $\alpha \in (0,\bar{\alpha})$ where $\bar{\alpha} \defeq \min\{\alpha_{1},\alpha_{b},\widehat{\alpha},\alpha_{s}\}$ and $\omega\in\Omega_1\cap\Omega_2$. When $\scriptS = \{0\}$, set $\alpha_{s} = \infty$. Then, we obtain the result of Theorem~\ref{thm:main}. From Lemma~\ref{lem:local_fast_part}, \eqref{eqn:T_estimate}, \eqref{eqn:scalar_recursion}, and \eqref{eqn:recursionforY}, $\gamma_{c}$ is given as 
    \vspace{-4pt}
    \begin{equation*}
        \gamma_{c} {\defeq} 
        \begin{cases}
            \max \{\gamma_{R},\,(1+c_A\alpha)\rho_{b}^{p_K/(2NB)},\,1-\mu\alpha\},& \scriptS {\neq} \{0\}\\
            \max \{\gamma_{R},\,(1+c_A\alpha)\rho_{b}^{p_K/(2NB)}\},& \scriptS {=} \{0\}.\\
        \end{cases}
    \end{equation*}

    \subsection{Parameter Selection}\label{sec:parameter}

        For convenience, set $\ell_{i} = \scriptL_{i} = \ell$ for all $i \in \scriptN$. Intuitively, larger $\ell$ would increase the probability of projection-averaging and may be expected to improve the observer performance. For $B>1$, however, the proof establishes the contraction through a particular event: each agent does not perform projection-averaging during the first $B-1$ steps and performs projection-averaging at the final step. Although increasing $\ell$ makes the projection-averaging step more likely, it makes the preceding $B-1$ steps less likely. Consequently, choosing $\ell$ too large decreases the probability of the particular event used in the proof. This can be observed from the form of $p_{K}$, which is maximized at $\ell^{\star}=1/(B-1)$. This results from the particular event used in the proof, and is not claimed to characterize the actual performance of the observer. For the effects of other parameters refer to Appendix~\ref{app:parameter}.

    \subsection{Comparison with \cite{KLS25}}\label{sec:KLS25}
    
        The observer proposed in this paper is closely related to that in \cite{KLS25}. However, \cite{KLS25} requires the commutativity condition $AP_{i}=P_{i}A$ for $i\in\scriptN$, whereas the present approach does not. The two analyses employ closely related decompositions that handle the observable part similarly (Lemma~\ref{lem:local_fast_part} and \cite[Prop. 1]{KLS25}), but handle the unobservable dynamics differently. In \cite{KLS25}, the transformation involving $e^{-At}$ removes the accumulated plant evolution. The commutativity condition $AP_{i}=P_{i}A$ keeps the transformed projectors fixed, since $e^{-At}P_{i} e^{At}=P_{i}$. Hence, the homogeneous part of \cite[Eq. (9)]{KLS25} falls exactly within the class covered by \cite[Th. 4]{LMM18}, enabling the contraction argument that underpins the proof.
        
        Our proof does not seek such cancellation. In Lemma~\ref{lem:homogeneous}, we have shown that even without the commutativity condition, the contraction argument holds for sufficiently small $\alpha \in \doubleR_{>0}$. The key idea of the proof is as follows: we first establish a contraction at $\alpha=0$, where the plant dynamics do not affect the update. The resulting projection-averaging structure is analogous to that obtained in \cite{KLS25} after canceling the plant evolution. We then show that, for sufficiently small $\alpha$, the actual update remains sufficiently close to the case $\alpha=0$, so that the contraction property is preserved. We trade the structural commutativity condition for a time scale restriction, requiring sufficiently small $\alpha$ to limit plant evolution per communication opportunity.

\section{Simulation}\label{sec:simulation}  
    \begin{figure}[t]
    \centering
    \begin{tikzpicture}[
        >=stealth,
        x=0.65cm,
        y=0.65cm,
        every node/.style={inner sep=1.2pt,font=\normalsize},
        agent/.style={
            circle,
            draw,
            minimum size=5.5mm,
            inner sep=0pt
        },
        comm/.style={->,thick,red!75!black},
        commoff/.style={->,thick,draw=gray!25}
    ]
        
        \newcommand{\drawagents}{
            \node[agent] (n1) at (0,0.8) {1};
            \node[agent] (n2) at (1.9,0.8) {2};
            \node[agent] (n3) at (1.9,-0.8) {3};
            \node[agent] (n4) at (0,-0.8) {4};
        }
        
        \begin{scope}[shift={(0,0)}]
            \drawagents
            
            \draw[commoff] (n1) -- (n2);
            \draw[commoff] (n2) -- (n3);
            \draw[commoff] (n3) -- (n4);
            \draw[commoff] (n4) -- (n1);
            
            \draw[comm] (n1) --
                node[midway,above=2pt,font=\scriptsize]
                {$p_{\mathrm{link}}$}
                (n2);
        
            \node[anchor=north,align=center,font=\normalsize]
                at (0.95,-1.35)
                {(a) 1 $\rightarrow$ 2};
        \end{scope}
        
        \begin{scope}[shift={(3.5,0)}]
            \drawagents
            
            \draw[commoff] (n1) -- (n2);
            \draw[commoff] (n2) -- (n3);
            \draw[commoff] (n3) -- (n4);
            \draw[commoff] (n4) -- (n1);
            
            \draw[comm] (n2) --
                node[midway,left=2pt,font=\scriptsize]
                {$p_{\mathrm{link}}$}
                (n3);
        
            \node[anchor=north,align=center,font=\normalsize]
                at (0.95,-1.35)
                {(b) 2 $\rightarrow$ 3};
        \end{scope}
        
        \begin{scope}[shift={(7.0,0)}]
            \drawagents
            
            \draw[commoff] (n1) -- (n2);
            \draw[commoff] (n2) -- (n3);
            \draw[commoff] (n3) -- (n4);
            \draw[commoff] (n4) -- (n1);
            
            \draw[comm] (n3) --
                node[midway,below=2pt,font=\scriptsize]
                {$p_{\mathrm{link}}$}
                (n4);
        
            \node[anchor=north,align=center,font=\normalsize]
                at (0.95,-1.35)
                {(c) 3 $\rightarrow$ 4};
        \end{scope}
        
        \begin{scope}[shift={(10.5,0)}]
            \drawagents
            
            \draw[commoff] (n1) -- (n2);
            \draw[commoff] (n2) -- (n3);
            \draw[commoff] (n3) -- (n4);
            \draw[commoff] (n4) -- (n1);
            
            \draw[comm] (n4) --
                node[midway,right=2pt,font=\scriptsize]
                {$p_{\mathrm{link}}$}
                (n1);
        
            \node[anchor=north,align=center,font=\normalsize]
                at (0.95,-1.35)
                {(d) 4 $\rightarrow$ 1};
        \end{scope}
        
    \end{tikzpicture}
    \vspace{-6mm}
    \caption{At the beginning of each $T_{r}$, one mode from (a)--(d) is selected independently and uniformly and is retained throughout the interval. Conditional on the past and the selected mode, its single edge is available independently at each iteration with probability $p_{\rm link}$.}
    \label{fig:schedule}
    \vspace{-6mm}
\end{figure}
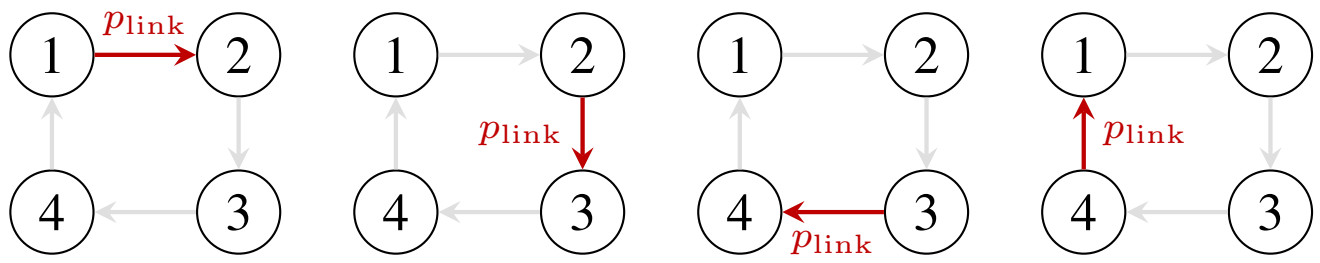
    \begin{figure*}[!t]
    \centering
    \includegraphics[width=\linewidth]{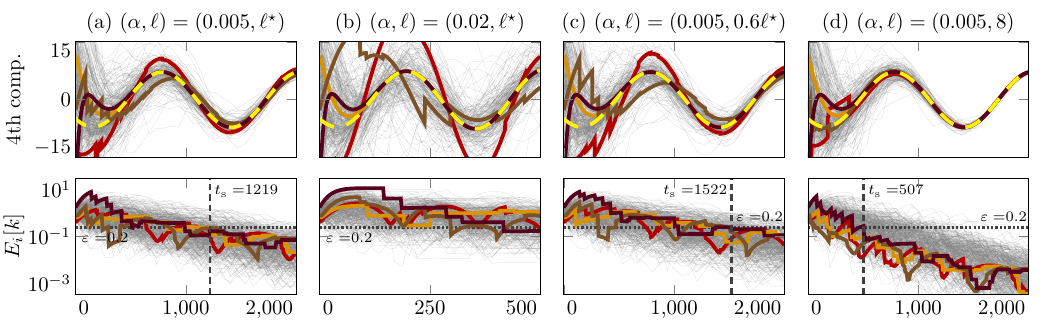}
    \vspace{-7mm}
    \caption{Trajectories of the fourth state component of each agent (top) and $E_{i}[k] \defeq \|e_{i}[k]\|/\max_{j}\|e_{j}[0]\|$ (bottom). The yellow dashed curves denote the true fourth state component of $x[k]$, the other colored curves correspond to the sampled trajectory, and the light-gray curves collect all $400$ trajectories from $100$ Monte Carlo simulations. The settling time $t_{\mathrm{s}}$ (with $\varepsilon = 0.2$) is marked by a vertical dashed line.}
    \label{fig:simulation}
    \vspace{-4mm}
\end{figure*}

    Consider $N=4$ agents and
    \vspace{-2pt}
    \begin{equation*}
        A=
        \begin{bmatrix}
            0&1&5&0\\-1&0&0&-5\\
            0&0&0&1\\0&0&-1&0
        \end{bmatrix},\quad
        C=
        \begin{bmatrix}
            1&0&1&0\\0&0&1&0\\
            0&1&0&1\\0&0&0&1
        \end{bmatrix},
        \vspace{-2pt}
    \end{equation*}
    where $C=\operatorname{col}(C_1,\ldots,C_4)$ as in Assumption~\ref{asm:jointobs}. The poles of each reduced observer are placed at $-2$ and $-3$. The pair $(C,A)$ is detectable, whereas every local pair $(C_{i},A)$ is not. Moreover, $\|AP_2-P_2A\|=\|AP_4-P_4A\|=5$, and hence the commutativity condition required in \cite{KLS25} is violated. Time is partitioned into mode intervals $T_{r} \defeq \{10r,\ldots,10r+9\}$. At the beginning of each $T_{r}$, one of the four modes in Figure~\ref{fig:schedule} is selected independently and uniformly, and the selected edge is available at each iteration of $T_{r}$ independently with probability $p_{\rm link}$, as shown in Figure~\ref{fig:schedule}. We take $B=31$, which is the smallest possible USSC horizon for this communication process. Indeed, every $31$ step block intersects four mode intervals and therefore has a uniformly positive conditional probability of containing all four communication modes specified in Figure~\ref{fig:schedule}. In contrast, the initial block of any length $B\leq30$ intersects at most three mode intervals, so its union graph can never be strongly connected. Fixing $B=31$ and denoting $\widehat{p}_{\mathrm{link}} \defeq 1 - p_{\mathrm{link}}$, a lower bound for the conditional probability of strong connectivity is computed as
    \vspace{-6pt}
    \begin{equation*}
        \begin{aligned}
            &\scriptP\!\left(
                \scriptG([31m,31m+30])
                \text{ is strongly connected}
                \,\middle|\,\scriptF_{31m}
            \right)
            \\
            &\quad\geq \min_{z \in \{0,\ldots,9\}}
            \frac{4!}{4^{4}}
            \left[1-\widehat{p}_{\mathrm{link}}^{10-z}\right]
            \left[1-\widehat{p}_{\mathrm{link}}^{10}\right]^{2}
            \left[1-\widehat{p}_{\mathrm{link}}^{z+1}\right]
            \\
            &\quad\geq
            \frac{3}{32}p_{\rm link}
            \left[1-\widehat{p}_{\mathrm{link}}^{10}\right]^{3} \eqdef q.
        \end{aligned}
        \vspace{-3pt}
    \end{equation*}
    Setting the above lower bound to $q=0.002$ and solving for $p_{\rm link}$ yields $p_{\rm link}\approx0.0913$, which is used in the simulations. Also, define the settling time $t_{\mathrm{s}}\defeq \inf\{k\in\{0,\ldots,k_{\max}\}:\max_{i}[\|e_{i}[t]\|/\max_{j}\|e_{j}[0]\|]\leq\varepsilon,\ \forall t=k,\ldots,k_{\max}\}$, where $k_{\max} \in \doubleN$ denotes the maximum iteration index. 
    
    In Figure~\ref{fig:simulation}, (a) and (b) compare $\alpha=0.005$ and $0.02$. The former achieves better tracking performance, compared with the latter case. This is in line with the role of $\alpha$ discussed after Theorem~\ref{thm:main}, which guarantees convergence for small $\alpha$. We also remark that the step sizes used here are much larger than the bound $\bar\alpha$ of Theorem~\ref{thm:main}, which is highly conservative owing to the particular event used in the proof (cf.\ Section~\ref{sec:parameter}); the simulations thus illustrate the qualitative roles of $\alpha$ and $\ell$ rather than the bound itself. (a), (c), and (d) compare $\ell=\ell^{\star}$, $0.6\ell^{\star}$, and $8$, respectively, at $\alpha=0.005$, where $\ell^{\star}=1/(B-1)=1/30$ as defined in Section~\ref{sec:parameter}. Although $p_{K}$ is maximized at $\ell^{\star}$ (cf. \ Section~\ref{sec:parameter}) (d) settles much faster than (a) and (c), suggesting that the nonmonotonic $\ell$-dependence induced by the proof event may be conservative. 

    \vspace{-5pt}

\section{Conclusion}\label{sec:conclusion}
    This paper proposed a distributed observer for forward-Euler implementations of continuous-time LTI systems under random local operations and random communication networks. Under joint detectability and the USSC condition, almost sure exponential convergence of all local state estimates toward the true state is achieved for sufficiently small step sizes. Possible directions for future work include developing continuous-time counterpart of this framework, incorporating renewal process based activation mechanisms regarding $\Delta_{i}^{k}$, and extending the framework to nonlinear settings.
    
\appendices


\section{Proof of Lemma~\ref{lem:local_fast_part}}\label{app:local_fastpart}
    
    Let $\bar e_{i}[k]\defeq M_{i}^{\top} e_{i}[k]$. Since $Q_{i}=M_{i}M_{i}^{\top}$, it is enough to study $\bar e_{i}[k]$. Since $\bar{\scriptO}_{i}$ is an invariant subspace of $A-L_{i}C_{i}$, $M_{i}^{\top}(A-L_{i}C_{i})P_{i} = 0$. Thus,
    \vspace{-4pt}
    \begin{equation*}
        \begin{aligned}
            M_{i}^{\top} e_{i}[k^{+}]
            &=
            M_{i}^{\top}
            \bigl(I_{n}+\alpha(A-L_{i}C_{i})\bigr)e_{i}[k]
            \\
            &=
            \bigl(I_{n-n_{i}}
            +
            \alpha(\bar A_{i}-\bar L_{i}\bar C_{i})\bigr)\bar e_{i}[k].
        \end{aligned}
        \vspace{-4pt}
    \end{equation*}
    Since $M_{i}^{\top} P_{i}=0$, we obtain
    \vspace{-4pt}
    \begin{equation*}
        \bar e_{i}[k+1]
        =
        \bigl(I_{n-n_{i}}
        +
        \alpha(\bar A_{i}-\bar L_{i}\bar C_{i})\bigr)\bar e_{i}[k].
        \vspace{-4pt}
    \end{equation*}
    Since $\bar A_{i}-\bar L_{i}\bar C_{i}$ is Hurwitz for every $i\in\scriptN$, we can choose 
    \vspace{-3pt}
    \begin{equation}\label{eqn:alpha_1}
        \alpha_1 \in (0,\min_{i \in \scriptN}\alpha_{\mathrm{E}}(\bar{A}_{i} - \bar{L}_{i}\bar{C}_{i})),
        \vspace{-3pt}
    \end{equation}
    where $\alpha_{\mathrm{E}}(\cdot)$ is the forward-Euler stability threshold defined in notation the paragraph of Section~\ref{sec:introduction}, such that $I_{n-n_{i}} + \alpha(\bar A_{i}-\bar L_{i}\bar C_{i})$ is Schur for every $i\in\scriptN$ and every $\alpha\in(0,\alpha_1)$ (with the convention that $\alpha_{\mathrm{E}}$ for a zero-dimensional matrix is $+\infty$). Since $\scriptN$ is finite, there exist deterministic constants $c_Q\in\doubleR_{>0}$ and $\gamma_Q\in(0,1)$ such that 1) of Lemma~\ref{lem:local_fast_part} holds. It is straightforward to show that $\gamma_{Q}$ is given as in \eqref{eqn:gamma_Q}. 

    We now prove the second assertion. Define the age of $e_{i|j}[k]$ by $a_{i|j}[k]\defeq k-\tau_{i|j}[k]$. Then, for every integer $r\in\{1,\ldots,k\}$,
    \vspace{-6pt}
    \begin{equation*}
        \{a_{i|j}[k]\geq r\}
        \subseteq
        \{\Delta_{i}^{k-r}=1,\ldots,\Delta_{i}^{k-1}=1\}.
        \vspace{-6pt}
    \end{equation*}
    Indeed, if a buffer error $e_{i|j}[k]$ has age at least $r$ and remains available at step $k$, no buffer-reset can have occurred during the preceding $r$ steps.
    By the tower property,
    \vspace{-4pt}
    \begin{align*}
        &\scriptP(\{\Delta_{i}^{k-r}=1,\ldots,\Delta_{i}^{k-1}=1\})
        \\
        &=\doubleE
        [\mathbf{1}_{\{\Delta_{i}^{k-r}=1,\ldots,\Delta_{i}^{k-1}=1\}}]\\ 
        &=
        \doubleE
        [\mathbf{1}_{\{\Delta_{i}^{k-r}=1,\ldots,\Delta_{i}^{k-2}=1\}}\doubleE[\mathbf{1}_{\{\Delta_{i}^{k-1}=1\}}|\scriptF_{k-1}]]
        \\
        &\leq
        \frac{1}{\ell_{i}+1}
        \doubleE[\mathbf{1}_{\{\Delta_{i}^{k-r}=1,\ldots,\Delta_{i}^{k-2}=1\}}] \leq \cdots \leq \left(\frac{1}{\ell_{i}+1}\right)^{r} .
        \vspace{-4pt}
    \end{align*}
    Denoting $\ell_{\min} \defeq \min_{i \in \scriptN}\ell_{i}$, choose $c_{a} \in \doubleR_{>0}$ satisfying $c_{a}\log(\ell_{\min}+1)>2$, and set $r_{k}\defeq \lceil c_{a}\log(k+2)\rceil$. Then, for all sufficiently large $k$ satisfying $r_{k} \leq k$,
    \vspace{-4pt}
    \begin{equation*}
        \begin{aligned}
            &\scriptP
            \Big(
                \max_{\substack{
                    i\in\scriptN\\
                    j\in\scriptM_{i}[k]\setminus\{i\}
                }}
                a_{i|j}[k]\geq r_k
            \Big)
            \\
            &\leq
            \sum_{i\in\scriptN}
            \sum_{j\in\scriptN\setminus\{i\}}
            \scriptP
            (
                j\in\scriptM_{i}[k],\
                a_{i|j}[k]\geq r_{k}
            )
            \\
            &\leq
            \sum_{i\in\scriptN}
            \sum_{j\in\scriptN\setminus\{i\}}
            \scriptP
            \Big(
                \bigcap_{s=k-r_{k}}^{k-1}
                \{\Delta_{i}^{s}=1\}
            \Big)
            \\
            &\leq
            (N-1)
            \sum_{i \in \scriptN}\left(\frac{1}{\ell_{i}+1}\right)^{r_{k}}
            \\
            &\leq
            N(N-1)
            \left(\frac{1}{\ell_{\min}+1}\right)^{r_{k}}.
        \end{aligned}
        \vspace{-4pt}
    \end{equation*}
    The right-hand side is summable over $k$. By the Borel--Cantelli lemma \cite[Th. 2.3.1]{Du19}, there exists an almost surely finite $k_{a}(\omega)\in\doubleN$ such that, for any $k \geq k_{a}(\omega)$,
    \vspace{-6pt}
    \begin{equation*}
        \max_{i\in\scriptN, j\in\scriptM_{i}[k]\setminus\{i\}}a_{i|j}[k] < r_{k},
        \vspace{-6pt}
    \end{equation*}
    where the maximum over an empty set is understood as zero. Denote $\Omega_{1} \defeq \{\omega \in \Omega:\exists k_{a}(\omega) < \infty\}$. Note that $\scriptP(\Omega_{1}) = 1$. Also, from the fact that $a_{i|j}[k]\leq k$, we obtain $a_{i|j}[k]\leq\max\{k_{a}(\omega),r_{k}\}$. Applying 1) of Lemma~\ref{lem:local_fast_part}, with the fact that
    $\|\Phi_{\alpha}^{r}\|\leq(1+\|A\|\alpha)^{r}$,
    \vspace{-4pt}
    \begin{align*}
        \|r_{i|j}[k]\|
        &{\leq}
        c_Q(1+\|A\|\alpha)^{k-\tau_{i|j}[k]-1}\max_{i\in\scriptN}\|Q_{i}e_{i}[0]\|\gamma_Q^{\tau_{i|j}[k]+1}
        \\
        &{=}
        c_Q\gamma_Q^{k}\max_{i\in\scriptN}\|Q_{i}e_{i}[0]\|\left(\frac{1+\|A\|\alpha}{\gamma_Q}\right)^{a_{i|j}[k]-1}
        \\
        &{\leq}
        c_{a}'(\omega)\gamma_Q^{k}
        (k{+}2)^{c_{a}\log((1{+}\|A\|\alpha)/\gamma_Q)}
        {\leq}
        c_R(\omega,\gamma_{R})\gamma_R^{k},
        \vspace{-4pt}
    \end{align*}
    for any $\gamma_R\in(\gamma_Q,1)$ and almost surely finite constants
    $c_{a}'(\omega),c_R(\omega,\gamma_{R}) \in \doubleR_{>0}$.

\section{Proof of Lemma~\ref{lem:unobservabledynamics}}\label{app:unobservabledynamics}

    Since $A\bar{\scriptO}_{i}\subset\bar{\scriptO}_{i}$, we have $P_{i}AP_{i}=AP_{i}$. Using the fact that $C_{i}P_{i}=0$ and $P_{i}L_{i}=0$, we obtain
    \begin{equation}\label{eqn:localprojectionid}
        \begin{aligned}
            u_{i}[k^{+}] 
            &\defeq P_{i} e_{i}[k^{+}]\\
            &=
            P_{i}
            \bigl(I_{n}+\alpha(A-L_{i}C_{i})\bigr)
            \bigl(P_{i} e_{i}[k]+Q_{i} e_{i}[k]\bigr)
            \\
            &=
            \Phi_{\alpha} u_{i}[k]
            +
            \alpha P_{i}AQ_{i} e_{i}[k].
        \end{aligned}
    \end{equation}
    Defining $\chi_{i}[k]\defeq\alpha P_{i}AQ_{i} e_{i}[k]$ and $\vartheta_{i}[k]\defeq Q_{i} e_{i}[k^{+}]$, we obtain from \eqref{eqn:localprojectionid} that $e_{i}[k^{+}]=\Phi_{\alpha} u_{i}[k]+\chi_{i}[k]+\vartheta_{i}[k]$.
    
    For $j\neq i$, exploiting \eqref{eqn:decomposition_detailed} and the fact that $P_{j}AP_{j} = AP_{j}$,
    \vspace{-5pt}
    \begin{equation*}
        u_{i|j}[k+1]
        =
        \begin{cases}
            \Phi_{\alpha} u_{j}[k]+\chi_{j}[k],
            & j\in\scriptM_{i}[k+1]\cap\scriptN_{i}[k],
            \\[1mm]
            \Phi_{\alpha} u_{i|j}[k],
            & j\in\scriptM_{i}[k+1]\setminus\scriptN_{i}[k],
            \\[1mm]
            0,
            & j\notin\scriptM_{i}[k+1].
        \end{cases}
        \vspace{-3pt}
    \end{equation*}
    Denoting $[\scriptT_{k}^{\alpha}\xi[k]]_{i|j}$ as the component of $\scriptT_{k}^{\alpha}\xi[k]$, corresponding to $v_{i|j}$ of $v$ in \eqref{eqn:stack}, it is straightforward to prove that 
    \vspace{-3pt}
    \begin{equation*} 
        [\scriptT_{k}^{\alpha}\xi[k]]_{i|j}
        = 
        \begin{cases} 
            \Phi_{\alpha} u_{j}[k], & j\in\scriptM_{i}[k+1]\cap\scriptN_{i}[k], \\[1mm] 
            \Phi_{\alpha} u_{i|j}[k], & j\in\scriptM_{i}[k+1]\setminus\scriptN_{i}[k], \\[1mm] 
            0, & j\notin\scriptM_{i}[k+1]. 
        \end{cases}
        \vspace{-3pt}
    \end{equation*} 
    Then, $u_{i|j}[k+1] = [\scriptT_{k}^{\alpha}\xi[k]]_{i|j} + d_{i|j}[k]$, where 
    \vspace{-3pt}
    \begin{equation*} 
        d_{i|j}[k] \defeq 
            \begin{cases} 
            \chi_{j}[k], & j\in\scriptM_{i}[k+1]\cap\scriptN_{i}[k], \\[1mm] 
            0, & \text{otherwise}. 
        \end{cases} 
        \vspace{-3pt}
    \end{equation*} 
    
    For $j\in\scriptM_{i}^{+}[k]$ (including $i$), write
    $e_{i|j}[k^{+}]=h_{i|j}[k]+\varepsilon_{i|j}[k]$, where
    \vspace{-4pt}
    \begin{align*}
        h_{i|j}[k]
        &\defeq
        \begin{cases}
            \Phi_{\alpha} u_{j}[k],
            & j\in \{i\} \cup \scriptN_{i}[k],
            \\[1mm]
            \Phi_{\alpha} u_{i|j}[k],
            & j\in\scriptM_{i}[k]\setminus(\scriptN_{i}[k]\cup\{i\}),
        \end{cases}
        \\
        \varepsilon_{i|j}[k]
        &\defeq
        \begin{cases}
            \chi_{j}[k]+\vartheta_{j}[k],
            &j \in \{i\}\cup\scriptN_{i}[k],
            \\[1mm]
            \Phi_{\alpha} r_{i|j}[k],
            & j\in\scriptM_{i}[k]\setminus(\scriptN_{i}[k]\cup\{i\}).
        \end{cases}
    \end{align*}
    For notational convenience, we use the notation $e_{i|i}[k^{+}]$, rather than $e_{i}[k^{+}]$, only in this passage. 
    Denoting $[\scriptT_{k}^{\alpha}\xi[k]]_{i}$ as the component of $\scriptT_{k}^{\alpha}\xi[k]$, corresponding to $v_{i}$ of $v$ in \eqref{eqn:stack},
    \vspace{-4pt}
    \begin{equation*}
        [\scriptT_{k}^{\alpha}\xi[k]]_{i}
        =
        \begin{cases}
            \displaystyle
            \frac{1}{|\scriptM_{i}^{+}[k]|}
            P_{i}\hspace{-5pt}
            \sum_{j\in\scriptM_{i}^{+}[k]}\hspace{-5pt} h_{i|j}[k],
            & \Delta_{i}^{k}=0,
            \\[3mm]
            \Phi_{\alpha} u_{i}[k],
            & \Delta_{i}^{k} = 1.
        \end{cases}
        \vspace{-4pt}
    \end{equation*}
    Thus, if $\Delta_{i}^{k}=0$, from the fact that $P_{i}Q_{i} = 0$,
    \vspace{-4pt}
    \begin{equation*}
        \begin{aligned}
            u_{i}[k+1]
            &=
            P_{i} e_{i}[k+1]
            =
            \frac{1}{|\scriptM_{i}^{+}[k]|}
            P_{i}
            \sum_{j\in\scriptM_{i}^{+}[k]} e_{i|j}[k^{+}]
            \\
            &=
            [\scriptT_{k}^{\alpha}\xi[k]]_{i}
            +
            \frac{1}{|\scriptM_{i}^{+}[k]|}
            P_{i}
            \sum_{j\in\scriptM_{i}^{+}[k]}\varepsilon_{i|j}[k].
        \end{aligned}
        \vspace{-4pt}
    \end{equation*}
    If the projection-averaging step is not performed, then
    $u_{i}[k+1]=P_{i} e_{i}[k^{+}]=\Phi_{\alpha} u_{i}[k]+\chi_{i}[k]$, whereas
    $[\scriptT_{k}^{\alpha}\xi[k]]_{i}=\Phi_{\alpha} u_{i}[k]$. Therefore, 
    \vspace{-4pt}
    \begin{align*}
        d_{i}[k]
        \defeq
        \begin{cases}
            \displaystyle
            \frac{1}{|\scriptM_{i}^{+}[k]|}
            P_{i}\hspace{-5pt}
            \sum_{j\in\scriptM_{i}^{+}[k]}\hspace{-5pt}
            \varepsilon_{i|j}[k],
            & \Delta_{i}^{k}=0,
            \\[3mm]
            \chi_{i}[k],
            & \Delta_{i}^{k}=1.
        \end{cases}
    \end{align*}
    Stacking $d_{i}[k]$ and $d_{i|j}[k]$ in the same order as $\xi[k]$, we obtain the form of \eqref{eqn:robust}. Moreover, $d[k]$ is $(k+1)$-structurally compatible. Indeed, $d_{i}[k]\in\bar{\scriptO}_{i}$ because it is either $\chi_{i}[k]\in\bar{\scriptO}_{i}$ or $d_{i}[k]$ lies in the image of $P_{i}$. Also, $d_{i|j}[k] = \chi_{j}[k] \in \bar{\scriptO}_{j}$ whenever $j \in \scriptM_{i}[k+1]\cap \scriptN_{i}[k]$, and $d_{i|j}[k]=0$ otherwise.

\section{Proof of Lemma~\ref{lem:homogeneous}}\label{prf:lem:homogeneous}
    
    First, we present a lemma required for proving Lemma~\ref{lem:homogeneous}:
            
    \begin{sublem}\label{lem:projection_product}
        Let $\scriptH[k]=(\scriptN,\scriptE[k])$ with $k = 1,\ldots,N-1$ be deterministic strongly connected directed graphs. Consider
        \vspace{-2pt}
        \begin{equation*}
            z_{i}[k+1]
            =
            P_{i}
            \sum_{j\in\{i\}\cup\scriptN_{i}[k]}
            a_{ij}^{k} z_{j}[k],
            \quad
            z_{i}[1]\in\bar{\scriptO}_{i},
            \vspace{-2pt}
        \end{equation*}
        where $\scriptN_{i}[k]$ denotes the in-neighbor set of agent $i$ in $\scriptH[k]$, the matrix $[a_{ij}^{k}] \in \doubleR^{N \times N}$ is row-stochastic, and satisfies $a_{ij}^{k} \geq 1/N$ if $j \in \{i\}\cup\scriptN_{i}[k]$ and $a_{ij}^{k} = 0$ otherwise. Then, there exists $\rho_{p}\in(0,1)$, dependent on $N$ and $\bar{\scriptO}_{i}$, such that
        \vspace{-3pt}
        \begin{equation*}
            \max_{i\in\scriptN}\|z_{i}[N]\|_{\scriptS} \leq \rho_{p} \max_{i\in\scriptN}\|z_{i}[1]\|_{\scriptS}.
            \vspace{-3pt}
        \end{equation*}
    \end{sublem}

    \vspace{-6pt}
    
    \begin{proof}
        Since $\scriptS\subseteq\bar{\scriptO}_{i}$, the projection $P_{i}$ is nonexpansive with respect to $\|\cdot\|_{\scriptS}$: $\|P_{i} z\|_{\scriptS} \leq \|z\|_{\scriptS}$ for all $z \in \doubleR^{n}$. Hence,
        \vspace{-3pt}
        \begin{equation}\label{eqn:nonexpansive}
            \max_{i\in\scriptN}\|z_{i}[k+1]\|_{\scriptS} \leq
            \max_{i\in\scriptN}\|z_{i}[k]\|_{\scriptS}.
            \vspace{-3pt}
        \end{equation}
        Suppose that the conclusion is false. Then, for each $m\in\doubleN$, there exist strongly connected graphs $\scriptH_{m}[k]=(\scriptN,\scriptE_{m}[k])$, $k=1,\ldots,N-1$, admissible weights $a_{ij}^{k,m}$, and initial vectors $z_{i}^{m}[1]\in\bar{\scriptO}_{i}$ such that $\max_{i\in\scriptN}\|z_{i}^{m}[1]\|_{\scriptS}=1$ and $\max_{i\in\scriptN}\|z_{i}^{m}[N]\|_{\scriptS}>1-1/m$. Since the quotient seminorm is invariant under adding elements of
        $\scriptS$ and $\scriptS \subseteq \bar{\scriptO}_{i}$, we may choose the initial vectors $z_{i}^{m}[1]$ so that
        $\Pi_{\scriptS}z_{i}^{m}[1]=0$ for all $i$ and $m$. Then, it is clear that the set $\{z \in \doubleR^{n}:\|z\|_{\scriptS} \leq 1,\Pi_{\scriptS}z=0\}$ is compact. Moreover, the set of admissible strongly connected graph topologies is finite. For each fixed graph sequence, row-stochasticity and the uniform positive lower bound on all nonzero weights imply that the vector obtained by concatenating $\{a_{ij}^{k}:i,j \in \scriptN,\;k=1,\ldots,N-1\}$ ranges over a compact subset of $\doubleR^{N^{2}(N-1)}$. Hence, after passing to a subsequence, we may assume that the graph topology is fixed and that both the weights and the initial vectors converge; the uniform positive lower bound on $a_{ij}^{k}$ ensures that the limiting weights preserve the fixed graph topology. The limiting trajectory therefore satisfies
        \vspace{-4pt}
        \begin{equation}\label{eqn:limittrajectory}
            \max_{i\in\scriptN}\|z_{i}[1]\|_{\scriptS}=1,
            \quad
            \max_{i\in\scriptN}\|z_{i}[N]\|_{\scriptS}=1.
            \vspace{-4pt}
        \end{equation}
        By \eqref{eqn:nonexpansive}, equality must hold at every intermediate step. Let $\bar{z}_{i}[k]\defeq (I_{n}-\Pi_{\scriptS})z_{i}[k]$, so that $\|z_{i}[k]\|_{\scriptS}=\|\bar{z}_{i}[k]\|$. Choose $i_{N}\in\scriptN$ such that
        $\|\bar{z}_{i_{N}}[N]\|=1$, and set $\bar{z}\defeq \bar{z}_{i_{N}}[N]$. Define
        \vspace{-4pt}
        \begin{equation*}
            \scriptI_{k} \defeq
            \left\{
                i\in\scriptN:
                (I_{n}-\Pi_{\scriptS})z_{i}[k]=\bar{z}
            \right\}.
            \vspace{-4pt}
        \end{equation*}
        By construction, $i_{N}\in\scriptI_{N}$, and hence $\scriptI_{N}\neq\emptyset$. We claim that $\scriptI_{k+1}\subseteq\scriptI_{k}$. More precisely, we claim that
        \vspace{-3pt}
        \begin{equation*}
            \bigcup_{i\in\scriptI_{k+1}}
            \left(\{i\}\cup\scriptN_{i}[k]\right)
            \subseteq
            \scriptI_{k}.
            \vspace{-3pt}
        \end{equation*}
        Indeed, fix $k\in\{1,\ldots,N-1\}$ and $i\in\scriptI_{k+1}$. Using \eqref{eqn:limittrajectory} and the fact that $P_{i}$ is nonexpansive with respect to $\|\cdot\|_{\scriptS}$,
        \vspace{-3pt}
        \begin{equation*}
            \begin{aligned}
                1
                &=
                \|z_{i}[k+1]\|_{\scriptS}
                =
                \Bigg\|
                    P_{i}
                    \sum_{j\in\{i\}\cup\scriptN_{i}[k]}
                    a_{ij}^{k} z_{j}[k]
                \Bigg\|_{\scriptS}
                \\
                &\leq
                \sum_{j\in\{i\}\cup\scriptN_{i}[k]}
                a_{ij}^{k}\|P_{i}z_{j}[k]\|_{\scriptS}
                \leq
                \sum_{j\in\{i\}\cup\scriptN_{i}[k]}
                a_{ij}^{k}\|z_{j}[k]\|_{\scriptS}
                \leq
                1.
            \end{aligned}
            \vspace{-3pt}
        \end{equation*}
        Thus, all inequalities above are equalities. Since $a_{ij}^{k}\in \doubleR_{>0}$ for every $j\in\{i\}\cup\scriptN_{i}[k]$, we have $\|P_{i} z_{j}[k]\|_{\scriptS} = \|z_{j}[k]\|_{\scriptS} = 1$ for all such $j$. Moreover, equality in the triangle inequality implies that the vectors $(I_{n}-\Pi_{\scriptS})P_{i} z_{j}[k]$, $j\in\{i\}\cup\scriptN_{i}[k]$, are positively collinear, i.e., there exist $c_{j}\in \doubleR_{\geq 0}$ such that
        \vspace{-3pt}
        \begin{equation*}
            (I_{n}-\Pi_{\scriptS})P_{i} z_{j}[k]
            =
            c_{j} (I_{n}-\Pi_{\scriptS})z_{i}[k+1],
            \;\;
            j\in\{i\}\cup\scriptN_{i}[k].
            \vspace{-3pt}
        \end{equation*}
        Since both sides have norm one, $c_{j}=1$. Therefore,
        \vspace{-3pt}
        \begin{equation*}
            (I_{n}-\Pi_{\scriptS})P_{i} z_{j}[k]
            =
            (I_{n}-\Pi_{\scriptS})z_{i}[k+1]
            =
            \bar{z},
            \;\;
            j\in\{i\}\cup\scriptN_{i}[k].
            \vspace{-3pt}
        \end{equation*}
        Since $\scriptS\subseteq\bar{\scriptO}_{i}$, the orthogonal projectors $P_{i}$ and $\Pi_{\scriptS}$ satisfy $(I_{n}-\Pi_{\scriptS})P_{i} = P_{i}(I_{n}-\Pi_{\scriptS})$. For $\|P_{i}z_{j}[k]\|_{\scriptS} = \|z_{j}[k]\|_{\scriptS}$, 
        \vspace{-3pt}
        \begin{equation*}
            \|P_{i}(I_{n}-\Pi_{\scriptS})z_{j}[k]\| = \|(I_{n}-\Pi_{\scriptS})z_{j}[k]\|.
            \vspace{-3pt}
        \end{equation*}
        Since $P_{i}$ is an orthogonal projector, this equality implies
        \vspace{-3pt}
        \begin{equation*}
            P_{i}(I_{n}-\Pi_{\scriptS})z_{j}[k]
            =
            (I_{n}-\Pi_{\scriptS})P_{i}z_{j}[k]
            =
            (I_{n}-\Pi_{\scriptS})z_{j}[k].
            \vspace{-3pt}
        \end{equation*}
        Consequently, we obtain
        \vspace{-3pt}
        \begin{equation*}
            (I_{n}-\Pi_{\scriptS})z_{j}[k]
            =
            (I_{n}-\Pi_{\scriptS})P_{i} z_{j}[k]
            =
            \bar{z},
            \quad
            j\in\{i\}\cup\scriptN_{i}[k].
            \vspace{-3pt}
        \end{equation*}
        Thus $j\in\scriptI_{k}$ for $j\in\{i\}\cup\scriptN_{i}[k]$, which proves the claim. 

        Now, we prove that $\scriptI_{1} = \scriptN$. It suffices to consider the case when $\scriptI_{N} \subsetneq \scriptN$. Whenever $\scriptI_{k+1}\subsetneq\scriptN$, the strong connectivity of $\scriptH[k]$ ensures that there exists an edge $(j,i)\in\scriptE[k]$ such that $i\in\scriptI_{k+1}$ and $j\notin\scriptI_{k+1}$. Since $j\in\scriptN_{i}[k]$, the claim gives $j\in\scriptI_{k}$. Hence, in this case, the inclusion is strict and $|\scriptI_{k}|\geq|\scriptI_{k+1}|+1$. Since $\scriptI_{N}$ is nonempty, applying this argument backward from $k=N-1$ to $k=1$ yields $\scriptI_{1}=\scriptN$. Therefore,
        \vspace{-3pt}
        \begin{equation*}
            (I_{n}-\Pi_{\scriptS})z_{i}[1]=\bar{z},
            \quad
            \forall i\in\scriptN.
            \vspace{-3pt}
        \end{equation*}
        Since $z_{i}[1]\in\bar{\scriptO}_{i}$ and $\scriptS\subseteq\bar{\scriptO}_{i}$, we have $\bar{z} = z_{i}[1]-\Pi_{\scriptS}z_{i}[1] \in \bar{\scriptO}_{i}$ for all $i \in \scriptN$. Thus, $\bar{z}\in\bigcap_{i=1}^{N}\bar{\scriptO}_{i}=\scriptS$. On the other hand, by construction, $\bar{z}\in\scriptS^{\perp}$ and $\|\bar{z}\|=1$, which is impossible. 
    \end{proof}

    \vspace{-5pt}

    Returning to the proof of Lemma~\ref{lem:homogeneous}, consider the homogeneous reduced dynamics \eqref{eqn:reduced_dynamics}, initialized at the step under consideration with an arbitrary structurally compatible state. Its local and buffer components are denoted by $\bar{u}_{i}[k]=\bar{u}_{i|i}[k]$ and $\bar{u}_{i|j}[k]$, respectively. We prove the contraction estimate in \eqref{eqn:alphacontraction}. Set
    \vspace{-3pt}
    \begin{equation*}
        Z[k]
        \defeq
        \max
        \left\{
            \|\bar{u}_{i}[k]\|_{\scriptS},\|\bar{u}_{i|j}[k]\|_{\scriptS}
            :
            i{\in}\scriptN,j{\in}\scriptM_{i}[k]\setminus \{i\}
        \right\}.
        \vspace{-3pt}
    \end{equation*}
    As in \eqref{eqn:T_estimate}, with $c_{A}$ defined in \eqref{eqn:c_A}, for every $\alpha\in\doubleR_{\geq0}$ and $k,T\in\doubleZ_{\geq0}$,
    \vspace{-4pt}
    \begin{equation}
        \begin{aligned}
            Z[k+T]
            &\leq
            (1+c_{A}\alpha)^{T} Z[k].
        \end{aligned}
        \vspace{-4pt}
    \end{equation}
    
    Suppose that the event $\scriptH_{m}$ occurs and $\scriptM_{i}[mB]=\{i\}$ for all $i\in\scriptN$. Define
    \vspace{-4pt}
    \begin{equation*}
        \scriptN_{i}^{m,B}
        \defeq
        \Bigg\{
            j\in\scriptN\setminus\{i\}:
            (j,i)\in
            \bigcup_{k=mB}^{(m+1)B-1}\scriptE[k]
        \Bigg\}.
            \vspace{-4pt}
    \end{equation*}
    Since $\Delta_{i}^{k}=1$ for $k=mB,\ldots,(m+1)B-2$, no projection-averaging step is performed before the last step $(m+1)B-1$. Hence, starting from $\scriptM_{i}[mB]=\{i\}$, the buffer of agent $i$ contains exactly the indices in $\{i\}\cup\scriptN_{i}^{m,B}$ when the final projection-averaging step is performed. Thus, when $\alpha = 0$, for all $i \in \scriptN$, 
    \vspace{-3pt}
    \begin{equation}\label{eqn:cleanupdate}
        \bar{u}_{i}[(m+1)B]
        =
        P_{i}
        \frac{1}{|\{i\}\cup\scriptN_{i}^{m,B}|}
        \sum_{j\in\{i\}\cup\scriptN_{i}^{m,B}}
        \hspace{-3pt}\bar{u}_{j}[mB].
        \vspace{-3pt}
    \end{equation}
    Since the graph with edge set $\{(j,i):j\in\scriptN_{i}^{m,B},\ i\in\scriptN\}$ is strongly connected on the event $\scriptH_{m}$ and $1/|\{i\}\cup\scriptN_{i}^{m,B}|$ is bounded below by $1/N$, \eqref{eqn:cleanupdate} is of the form in Lemma~\ref{lem:projection_product}.
    
    Now, suppose the event $\scriptK_{m}$ occurs. In the first event $\scriptH_{mN}$, it is not guaranteed that $\scriptM_{i}[mNB] = \{i\}$ for all $i \in \scriptN$. Thus, we use the first event only to flush those buffers; after this, $\scriptM_{i}[(mN+1)B]=\{i\}$ for all $i\in\scriptN$. The remaining events $\scriptH_{mN+1},\ldots,\scriptH_{(m+1)N-1}$ satisfy the conditions described above, and all buffers are empty at the endpoints of these blocks. Applying Lemma~\ref{lem:projection_product} at $\alpha=0$,
    \vspace{-4pt}
    \begin{equation}\label{eqn:cleancontraction}
        Z[(m+1)NB]
        \leq
        \rho_{p} Z[(mN+1)B],\quad \rho_{p} \in (0,1).
        \vspace{-4pt}
    \end{equation}
    In the present setting, $a_{ij}^{k}$ in Lemma~\ref{lem:projection_product} are uniquely determined by $1/|\{i\}\cup\scriptN_{i}^{s,B}|$. Since the set of strongly connected graph topologies is finite, $\rho_{p}$ can be chosen uniformly over all such graph sequences, with $\rho_{p}\in(0,1)$. 

    We now consider the case when $\alpha\in \doubleR_{>0}$. The initial buffer-index set $\scriptM_{i}[mNB]$ of event $\scriptH_{mN}$ may contain indices $j \neq i$ corresponding to copies created before time $mNB$. Note from the previous reasoning that the communication-refresh, buffer-reset, and projection-averaging do not increase $Z[k]$, while each application of $\Phi_{\alpha}$ increases the quotient seminorm by at most the factor $(1+c_{A}\alpha)$. Therefore, 
    \vspace{-4pt}
    \begin{equation}\label{eqn:cleancontraction_2}
        Z[(mN+1)B]
        \leq
        (1+c_{A}\alpha)^{B} Z[mNB].
        \vspace{-4pt}
    \end{equation}
    
    For the remaining events $\scriptH_{mN+1},{\ldots},\scriptH_{(m+1)N-1}$, it remains to justify that the contraction estimate in \eqref{eqn:cleancontraction} persists for sufficiently small $\alpha\in \doubleR_{>0}$. Under event $\scriptK_{m}$, \eqref{eqn:cleancontraction} can be represented as 
    \vspace{-5pt}
    \begin{equation*}
        \|
            \frakturB_{(m+1)N-1}^{0}
            \cdots
            \frakturB_{mN+1}^{0} \mathbf{z}
        \|_{\scriptS,\max}
        \leq
        \rho_{p}\|\mathbf{z}\|_{\scriptS,\max}
        \vspace{-4pt}
    \end{equation*}
    for every $\mathbf{z}=\operatorname{col}(z_{1},\ldots,z_{N})$ with $z_{i}\in\bar{\scriptO}_{i}$, by initializing the homogeneous dynamics at time $(mN+1)B$ with $\xi_{\mathbf{z}}$. The identity
    \vspace{-4pt}
    \begin{equation*}
        \begin{aligned}
            (I_{n}-\Pi_{\scriptS})\Phi_{\alpha}
            &=
            (I_{n}-\Pi_{\scriptS})(I_{n}+\alpha A)
            \bigl((I_{n}-\Pi_{\scriptS})+\Pi_{\scriptS}\bigr)
            \\
            &=
            (I_{n}-\Pi_{\scriptS})
            +
            \alpha(I_{n}-\Pi_{\scriptS})A(I_{n}-\Pi_{\scriptS})
            \\
            &=
            \left(
                I_{n}+\alpha(I_{n}-\Pi_{\scriptS})A(I_{n}-\Pi_{\scriptS})
            \right)
            (I_{n}-\Pi_{\scriptS}),
        \end{aligned}
        \vspace{-4pt}
    \end{equation*}
    implies that for every $k = 0,\ldots,B$, 
    \vspace{-4pt}
    \begin{equation*}
        \begin{aligned}
            &\|(\Phi_{\alpha}^{k}-I_{n})z\|_{\scriptS}\\
            &= \|(I_{n}-\Pi_{\scriptS})(\Phi_{\alpha}^{k}-I_{n})z\|\\
            &= \|([I_{n}+\alpha(I_{n}-\Pi_{\scriptS})A(I_{n}-\Pi_{\scriptS})
            ]^{k}-I_{n})(I_{n}-\Pi_{\scriptS})z\|\\
            &\leq
            \delta_{B}(\alpha)\|z\|_{\scriptS},
            \vspace{-9pt}
        \end{aligned}
    \end{equation*}
    where $\delta_{B}(\alpha)\defeq(1+c_{A}\alpha)^{B}-1 \in \doubleR_{\geq 0}$, with 
    \vspace{-3pt}
    \begin{equation*}
        c_{A} \defeq \|(I_{n}-\Pi_{\scriptS})A(I_{n}-\Pi_{\scriptS})\| \in \doubleR_{\geq 0}
        \vspace{-3pt}
    \end{equation*}
    (Note that $c_{A}$ is also defined in \eqref{eqn:c_A}; we repeat its definition here for ease of reading.) Then, for every $\mathbf{z}=\operatorname{col}(z_{1},\ldots,z_{N})$ with $z_{i}\in\bar{\scriptO}_{i}$ and every $s\in\{mN+1,\ldots,(m+1)N-1\}$, we claim that
    \vspace{-4pt}
    \begin{equation}\label{eqn:frakturBalpha}
        \|\frakturB_{s}^{\alpha} \mathbf{z}-\frakturB_{s}^{0} \mathbf{z}\|_{\scriptS,\max}
        \leq
        \delta_{B}(\alpha)\|\mathbf{z}\|_{\scriptS,\max}.
        \vspace{-4pt}
    \end{equation}
    Indeed, on event $\scriptH_{s}$, no projection-averaging step is performed before the last step. Starting with $\xi_{\mathbf{z}}$ at time $sB$, the local component of agent $j$ and every available buffer copy originating from it therefore equal $\Phi_{\alpha}^{B}z_{j}$ at the final intermediate stage, regardless of the refresh time. Thus, the $i$th local component of
    $\frakturB_{s}^{\alpha} \mathbf{z}-\frakturB_{s}^{0}\mathbf{z}$, denoted by $[\frakturB_{s}^{\alpha} \mathbf{z}-\frakturB_{s}^{0}\mathbf{z}]_{i}$, is given by
    \vspace{-3pt}
    \begin{equation*}
        [\frakturB_{s}^{\alpha} \mathbf{z}-\frakturB_{s}^{0}\mathbf{z}]_{i}
        =
        \frac{1}{|\{i\}\cup\scriptN_{i}^{s,B}|}
        \sum_{j\in\{i\}\cup\scriptN_{i}^{s,B}}P_{i}
        (\Phi_{\alpha}^{B}-I_{n})z_{j}.
        \vspace{-3pt}
    \end{equation*}
    Since $P_{i}$ is nonexpansive with respect to $\|\cdot\|_{\scriptS}$,
    \vspace{-3pt}
    \begin{equation*}
        \begin{aligned}
            \|[\frakturB_{s}^{\alpha} \mathbf{z}-\frakturB_{s}^{0}\mathbf{z}]_{i}\|_{\scriptS}
            &\leq
            \frac{1}{|\{i\}\cup\scriptN_{i}^{s,B}|}\hspace{-3pt}
            \sum_{j\in\{i\}\cup\scriptN_{i}^{s,B}}\hspace{-5pt}
            \|(\Phi_{\alpha}^{B}{-}I_{n})z_{j}\|_{\scriptS}
            \\
            &\leq
            \delta_{B}(\alpha)
            \max_{j\in\scriptN}\|z_{j}\|_{\scriptS}
            =
            \delta_{B}(\alpha)\|\mathbf{z}\|_{\scriptS,\max}.
        \end{aligned}
        \vspace{-3pt}
    \end{equation*}
    Taking the maximum over $i\in\scriptN$ gives \eqref{eqn:frakturBalpha}. Moreover, \eqref{eqn:nonexpansive} shows that $\frakturB_{s}^{0}$ is nonexpansive with respect to
    $\|\cdot\|_{\scriptS,\max}$, and hence \eqref{eqn:frakturBalpha} yields
    \vspace{-5pt}
    \begin{equation*}
        \|\frakturB_{s}^{\alpha} \mathbf{z}\|_{\scriptS,\max}
        \leq
        (1+\delta_{B}(\alpha))\|\mathbf{z}\|_{\scriptS,\max}.
        \vspace{-5pt}
    \end{equation*}
    Therefore, using a simple telescoping expansion, we obtain
    \vspace{-3pt}
    \begin{equation*}
        \begin{aligned}
            &\|
                \frakturB_{(m+1)N-1}^{\alpha}
                \cdots
                \frakturB_{mN+1}^{\alpha}\mathbf{z}
            \|_{\scriptS,\max}
            \\
            &\leq
            \|
                \frakturB_{(m+1)N-1}^{0}
                \cdots
                \frakturB_{mN+1}^{0}\mathbf{z}
            \|_{\scriptS,\max}\\
            &\quad +
            ((1+\delta_{B}(\alpha))^{N-1}-1)
            \|\mathbf{z}\|_{\scriptS,\max}
            \\
            &\leq
            \left[
                \rho_{p}
                +
                (1+\delta_{B}(\alpha))^{N-1}-1
            \right]
            \|\mathbf{z}\|_{\scriptS,\max}.
        \end{aligned}
        \vspace{-2pt}
    \end{equation*}
    Since $\delta_{B}(\alpha)\to0$ as $\alpha\to0$, fix $\rho_{b}\in(\rho_{p},1)$ and choose a deterministic $\alpha_{b}\in \doubleR_{>0}$, independent of $m$ and $\omega$, so that 
    \vspace{-3pt}
    \begin{equation}\label{eqn:alpha_b}
        (1+\delta_{B}(\alpha))^{N-1}-1\leq\rho_{b}-\rho_{p},\quad \alpha\in(0,\alpha_{b}).
        \vspace{-3pt}
    \end{equation}
    Then, for every $\alpha\in(0,\alpha_{b})$,
    \vspace{-3pt}
    \begin{equation*}
        \|
            \frakturB_{(m+1)N-1}^{\alpha}
            \cdots
            \frakturB_{mN+1}^{\alpha}\mathbf{z}
        \|_{\scriptS,\max}
        \leq
        \rho_{b}\|\mathbf{z}\|_{\scriptS,\max},
        \vspace{-3pt}
    \end{equation*}
    which concludes the proof.

\section{Parameter Selection \& Convergence Rate}\label{app:parameter}

    We focus on the homogeneous case $\ell_{i}= \mathscr{L}_{i} = \ell$ for all $i\in\scriptN$ and assume $\scriptS \neq \{0\}$. In this case, $\widehat{\alpha}$ in \eqref{eqn:widealpha} is given as $\widehat\alpha\defeq (\rho_{b}^{-p_K(\ell)/(2NB)}-1)/c_{A}$, where $p_{K}(\ell) \defeq q^N\ell^{N^{2}}/(\ell+1)^{N^{2}B}$ is the homogeneous version of $p_{K}$ in \eqref{eqn:pk}. From the proof of Theorem~\ref{thm:main}, when $\alpha \in (0,\min\{\alpha_{1},\alpha_{b},\widehat\alpha,\alpha_{s}\})$, the result holds almost surely, with 
    \vspace{-3pt}
    \begin{equation*}
        \gamma_{c} = \max\{\gamma_{R},(1+c_A\alpha)\rho_{b}^{p_K(\ell)/(2NB)},1-\mu\alpha\}.
        \vspace{-3pt}
    \end{equation*} 
    \begin{table}[t]
        \vspace{2.4mm}
        \centering
        \caption{Dependence of the step size and convergence rate bounds \vspace{-2mm}}
        \label{tab:rate_summary}
        \renewcommand{\arraystretch}{1.25}
        \setlength{\tabcolsep}{3pt}
        \footnotesize
        \begin{tabularx}{\columnwidth}{
            @{}
            >{\centering\arraybackslash}p{0.10\columnwidth}
            >{\raggedright\arraybackslash}X
            @{}
        }
            \hline
            Status
            &
            Sufficient conditions
            \\
            \hline
            \\[-10pt]
            $\alpha_1\uparrow$
            &
            $\displaystyle\min_{i\in\scriptN}\alpha_{\mathrm E}(\bar A_{i}-\bar L_{i}\bar C_{i})\uparrow$
            \\
            $\alpha_{s}\uparrow$
            &
            $\frakturS_{\max}(H_{\scriptS})\|H_{\scriptS}^{1/2}A|_{\scriptS}H_{\scriptS}^{-1/2}\|^{2}\downarrow$
            \\
            $\alpha_{b}\uparrow$
            &
            $c_A\downarrow$, $B(N-1)\downarrow$, or $\rho_{b}-\rho_{p}\uparrow$
            \\
            $\widehat{\alpha}\hspace{3pt}\uparrow$
            &
            $c_A\downarrow$, $\rho_{b}\downarrow$, or $p_K(\ell)/(NB)\uparrow$
            \\
            $\gamma_{c}\downarrow$
            &
            $\gamma_{R}\downarrow$, $(1+c_A\alpha)\rho_{b}^{p_K(\ell)/(2NB)}\downarrow$, and $\frakturS_{\max}(H_{\scriptS})\downarrow$ 
            \\[1pt]
            \hline
        \end{tabularx}
        \vspace{-4mm}
    \end{table}
    Dependencies are summarized in Table~\ref{tab:rate_summary}. (Dependencies on $\alpha_{1}$ and $\alpha_{b}$ are shown in \eqref{eqn:alpha_1} of Appendix~\ref{app:local_fastpart} and \eqref{eqn:alpha_b} of Appendix~\ref{prf:lem:homogeneous}, respectively.)
    \begin{itemize}[leftmargin=12pt]
        \item \textbf{(Subspace $\scriptS$):} Set two $A$-invariant subspaces $\scriptS_{1}$ and $\scriptS_{2}$, satisfying $\{0\} \subsetneq \scriptS_{1} \subseteq \scriptS_{2}$, and assume that $A|_{\scriptS_{2}}$ is Hurwitz. Then, from the definition of $c_{A}$ in \eqref{eqn:c_A},
            \vspace{-3pt}
            \begin{equation*}
                \|(I_{n} - \Pi_{\scriptS_{2}})A(I_{n} - \Pi_{\scriptS_{2}})\| \leq \|(I_{n} - \Pi_{\scriptS_{1}})A(I_{n} - \Pi_{\scriptS_{1}})\|,
                \vspace{-3pt}
            \end{equation*}
            roughly implying that a larger $\scriptS$ generally means smaller $c_{A}$, which improves $\alpha_{b}$, $\widehat{\alpha}$, and $\gamma_{c}$. Now, we note that 
            \vspace{-3pt}
            \begin{equation*}
                H_{\scriptS} = \int_{0}^{\infty} e^{(A|_{\scriptS})^{\top} t}e^{(A|_{\scriptS})t}dt.
                \vspace{-3pt}
            \end{equation*}
            For $\frakturS_{\max}(H_{\scriptS}) = \sup_{s \in \scriptS,\|s\| = 1}s^{\top}H_{\scriptS}s$, we deduce that $\frakturS_{\max}(H_{\scriptS_{1}}) \leq \frakturS_{\max}(H_{\scriptS_{2}})$. Also, we observe that
            \begin{equation*}
                \begin{aligned}
                    &\|H_{\scriptS_1}^{1/2}(A|_{\scriptS_1})H_{\scriptS_1}^{-1/2}\|
                    \\
                    &=\sup_{s\in\scriptS_1\setminus\{0\}}\frac{\|H_{\scriptS_1}^{1/2}(A|_{\scriptS_1})s\|}{\|H_{\scriptS_1}^{1/2}s\|}
                    =
                    \sup_{s\in\scriptS_1\setminus\{0\}}
                    \frac{\|H_{\scriptS_2}^{1/2}(A|_{\scriptS_2})s\|}{\|H_{\scriptS_2}^{1/2}s\|}
                    \\
                    &\leq\sup_{s\in\scriptS_2\setminus\{0\}}\frac{\|H_{\scriptS_2}^{1/2}(A|_{\scriptS_2})s\|}{\|H_{\scriptS_2}^{1/2}s\|}
                    = \|H_{\scriptS_2}^{1/2}(A|_{\scriptS_2})H_{\scriptS_2}^{-1/2}\|.
                \end{aligned}
            \end{equation*}
            Thus, larger $\scriptS$ deteriorates $\alpha_{s}$. For $\gamma_{c}$, the improvement due to a smaller $c_{A}$ competes with the deterioration due to a larger $\frakturS_{\max}(H_{\scriptS})$.

        \item \textbf{(Spectrum of $\bar A_{i}-\bar L_{i}\bar C_{i}$):} Placing the observer poles farther into the left-half plane may improve the continuous-time observer response. However, for a fixed $\alpha$, larger pole magnitudes can reduce $\alpha_{\mathrm E}(\bar A_{i}-\bar L_{i}\bar C_{i})$, thereby decreasing $\alpha_1$. Thus, excessively fast observer dynamics are not necessarily advantageous after Euler discretization.
    
    \end{itemize}

\bibliographystyle{ieeetr}
\bibliography{yongjoo}

\end{document}